\documentclass[12pt]{article}

\usepackage{cancel}
\usepackage{enumitem}
\usepackage{amssymb}
\usepackage{soul}
\usepackage{tabularx}
\usepackage{xcolor}
\usepackage{booktabs}
\usepackage{subcaption} 
\usepackage{colortbl}
\usepackage{authblk}
\usepackage{xspace}
\DeclareUnicodeCharacter{00A0}{ }

\newcolumntype{C}{>{\centering\arraybackslash}X}
\usepackage[T1]{fontenc}
\usepackage{epsfig}
\usepackage{latexsym}
\usepackage{graphicx}
\usepackage{amsmath}
\usepackage{amsfonts}   
\usepackage{amssymb}    
\usepackage{float}
\usepackage{bm}
\usepackage{hyperref} 
\usepackage[nodisplayskipstretch]{setspace}
\usepackage{url}
\def\lsim{\raise0.3ex\hbox{$\;<$\kern-0.75em\raise-1.1ex\hbox{$\sim\;$}}}
\def\gsim{\raise0.3ex\hbox{$\;>$\kern-0.75em\raise-1.1ex\hbox{$\sim\;$}}}

\def    \beq            {\begin{equation}}
\def    \eeq            {\end{equation}}
\def    \bea           {\begin{eqnarray}}
\def    \eea           {\end{eqnarray}}

\def \mn{\mu\nu{\rm SSM}}

\def\g2{{\rm GeV}^2}

\def\sw2{sin^2 \theta_w}

\def\a^tau{\alpha_{\tau}}

\def\beq{\begin{equation}}
\def\eeq{\end{equation}}
\def\beqa{\begin{eqnarray}}
\def\eeqa{\end{eqnarray}}

\newcommand{\tev}{\;\textrm{TeV}}
\newcommand{\gev}{\;\textrm{GeV}}

\newcommand{\newc}{\newcommand}
\newc\BR{BR}
\newc{\akappa}{A_{\kappa} }
\newc\deltagmtwo{\delta (g-2)_{\mu}} 
\newc\deltaamu{\Delta a_{\mu}}

\def\anti{\overline}

\def\la{\lambda}

\def\ka{\kappa}

\newc{\haa}{BR\(h_1\to a_1 a_1\)}
\newc{\abb}{BR\(a_1\to b\anti{b}\)}
\newc{\hbb}{BR\(h_1\to b\anti{b}\)}
\newc{\abund}{\Omega h^2}
\newc\bsgamma{b\rightarrow s \gamma }
\newc\bxsgamma{\overline{B}\rightarrow X_{s}\gamma}
\newc\brbsgamma{\BR(\overline{B}\rightarrow X_s\gamma)}

\newcommand{\fnalcen}{{204.0}}    
\newcommand{\fnalunc}{{5.4}}      
\newcommand{\fnalbnlsig}{{0.8}}   
\newcommand{\newcen}{{206.1}}     
\newcommand{\newunc}{{4.1}}       
\newcommand{\newdiff}{{25.1}}    
\newcommand{\newdiffunc}{{5.9}}   
\newcommand{\newdiffsig}{{4.2}}   

\newcommand\ReDiag{\mathop{%
  \raise .5pt\hbox{[}%
  \widetilde{\mathrm{Re}}%
  \raise .5pt\hbox{]}}}
\newcommand\ReOffDiag{\mathop{%
  \raise .5pt\hbox{$\llbracket$}%
  \widetilde{\mathrm{Re}}%
  \raise .5pt\hbox{$\rrbracket$}}}

\newcommand\refeq[1]{Eq.~(\ref{#1})}

\newcommand\refse[1]{Sect.~\ref{#1}}

\newcommand\citere[1]{Ref.~\cite{#1}}
\newcommand\citeres[1]{Refs.~\cite{#1}}

\newcommand{\mnSSM}{\ensuremath{\mu\nu\mathrm{SSM}}}

\newcommand\mmuesneu{m_{\tilde{\nu}_{\mu}}}

\newcommand{\sig}{\sigma}

\def\order#1{\ensuremath{{\cal O}(#1)}}
\def\reffi#1{\mbox{Fig.~\ref{#1}}}

\def\De{\Delta}

\def\la{\lambda}

\def\gmin2{\ensuremath{(g-2)_\mu}}
\def\amu{\ensuremath{a_\mu}}

\definecolor{Orange}{named}{orange}
\definecolor{Purple}{named}{purple}
\definecolor{Lightblue}{cmyk}{0.9,0.1,0.1,0.3}
\definecolor{dgelborange}{cmyk}{0.,0.3,0.5, 0.}
\definecolor{Lila}{rgb}{0.5,0.,1}
\definecolor{Darkgreen}{rgb}{0.,.7,0.2}

\allowdisplaybreaks

\usepackage{array, makecell}
\usepackage{boldline}
\usepackage[nosort]{cite}

\title{\bf{
The new {\boldmath{$(g-2)_\mu$}} result and the \boldmath{$\mu\nu$SSM}} 
}

\author[a,b,c]{Sven Heinemeyer\thanks{Sven.Heinemeyer@cern.ch}}
\author[a,d]{Essodjolo Kpatcha\thanks{kpatcha.essodjolo@uam.es}}
\author[e]{I\~naki Lara\thanks{inaki.lara@fuw.edu.pl}}
\author[f,g]{Daniel~E.~L\'opez-Fogliani\thanks{daniel.lopez@df.uba.ar}}
\author[a,d]{Carlos~Mu\~noz\thanks{c.munoz@uam.es}} 
\author[h]{Natsumi Nagata\thanks{natsumi@hep-th.phys.s.u-tokyo.ac.jp}}

\affil[a]{Instituto de F\'{\i}sica Te\'{o}rica (IFT) UAM-CSIC, 
  Campus de Cantoblanco, 28049 Madrid, Spain}
\affil[b]{Campus of International Excellence UAM+CSIC, 
Cantoblanco, 28049, Madrid, Spain}
\affil[c]{Instituto de F\'isica de Cantabria (CSIC-UC), 
39005, Santander, Spain}
\affil[d]{Departamento de F\'{\i}sica Te\'{o}rica, Universidad Aut\'{o}noma de Madrid (UAM),
Campus de Cantoblanco, 28049 Madrid, Spain}
 \affil[e] 
  {Faculty of Physics, University of Warsaw, Pasteura 5, 02-093 Warsaw, Poland}
  \affil[f]{Instituto de F\'isica de Buenos Aires UBA \& CONICET, Departamento de F\'isica,
 Facultad de Ciencia Exactas y Naturales, Universidad de Buenos Aires, 
1428 Buenos Aires, Argentina}
\affil[g]{
{Pontificia Universidad Católica Argentina, 
Av. Alicia Moreau de Justo 1500, 
1107~Buenos~Aires, Argentina}}
\affil[h] {Department of Physics, University of Tokyo, 
Tokyo 113-0033, Japan}

\date{}

\begin{document}

\maketitle

\begin{abstract}
The \mnSSM\ is a highly predictive alternative model to the MSSM. In particular, the
electroweak sector of the model can explain the longstanding
discrepancy between the experimental result for the anomalous magnetic
moment of the muon, \gmin2, and its Standard Model prediction,
while being in agreement with all other theoretical and experimental
constraints. 
The recently published MUON G-2 result is within
$\fnalbnlsig\,\sig$ in agreement with  the older BNL result on \gmin2.
The combined result was announced as 
$\amu^{\rm exp} = (11 659 \newcen \pm \newunc) \times 10^{-10}$,
yielding a new deviation from the Standard Model prediction of
$\De\amu = (\newdiff \pm \newdiffunc) \times 10^{-10}$, corresponding to
$\newdiffsig\,\sig$. Using this improved bound we update the analysis in the \mnSSM\ as
presented in Ref.~\cite{Kpatcha:2019pve} and set new limits on the allowed
parameters space of the electroweak sector of the model.
We conclude that significant regions of the model can explain the new \gmin2\ data.
\end{abstract}


\begin{flushright}
  IFT-UAM/CSIC-21-034
\end{flushright}

\clearpage 


\section{Introduction}
\label{sec:intro}

The experimental result for the anomalous magnetic moment of the muon, 
$\amu := \frac{1}{2}$\,\gmin2, 
was so far dominated by the measurements made at Brookhaven National Laboratory 
(BNL)~\cite{Bennett:2006fi},
resulting in a world average of~\cite{Tanabashi:2018oca}
\begin{align}
\amu^{\rm exp-BNL} &= (11 659 209.1 \pm 6.3) \times 10^{-10}~,
\label{gmt-exp-BNL}
\end{align}
combining statistical and systematic uncertainties.
The Standard Model (SM) prediction of \amu\ is given by~\cite{Aoyama:2020ynm}
(based on \citeres
{Keshavarzi:2019abf,Davier:2019can,Aoyama:2012wj,Aoyama:2019ryr,Czarnecki:2002nt,Gnendiger:2013pva,Davier:2017zfy,Keshavarzi:2018mgv,Colangelo:2018mtw,Hoferichter:2019mqg,Kurz:2014wya,Melnikov:2003xd,Masjuan:2017tvw,Colangelo:2017fiz,Hoferichter:2018kwz,Gerardin:2019vio,Bijnens:2019ghy,Colangelo:2019uex,Blum:2019ugy,Colangelo:2014qya}),
%
\begin{align}
\amu^{\rm SM} &= (11 659 181.0 \pm 4.3) \times 10^{-10}~, 
\label{gmt-sm}
\end{align}
corresponding to a $3.7\,\sig$ discrepancy.


Recently, the MUON G-2 collaboration~\cite{Grange:2015fou}
published the results (referred to as ``FNAL'' result) of their Run~1 data~\cite{Abi:2021gix,Albahri:2021ixb}, 
\begin{align}
\amu^{\rm exp-FNAL} &= (11 659 \fnalcen \pm \fnalunc) \times 10^{-10}~,
\label{gmt-exp-FNAL}
\end{align}
being within $\fnalbnlsig\,\sig$ well compatible with the previous
experimental result in \refeq{gmt-exp-BNL}.
The combined experimental result was announced as
\begin{align}
\amu^{\rm exp} &= (11 659 \newcen \pm \newunc) \times 10^{-10}~.
\label{gmt-exp}
\end{align}
Compared with the SM prediction in \refeq{gmt-sm}, this yields a new
deviation of
\begin{align}
\Delta\amu &= (\newdiff \pm \newdiffunc) \times 10^{-10}~, 
\label{gmt-diff}
\end{align}
corresponding to a $\newdiffsig\,\sig$ discrepancy.\footnote{A recent calculation in lattice QCD of the contribution of the hadronic vacuum 
polarization~\cite{Borsanyi:2020mff}, results in that there is no essential discrepancy with the experimental data. However, the authors of
Ref.~\cite{Lehner:2020crt} have argued that the estimation of the uncertainty of the lattice calculation is too optimistic. In addition, several authors have pointed out that such a result implies a tension with electroweak 
fits~\cite{Crivellin:2020zul,Keshavarzi:2020bfy,deRafael:2020uif}.}

In Ref.~\cite{Kpatcha:2019pve}, some of us performed an analysis of 
the (then current) deviation of \amu~\cite{Keshavarzi:2019abf,Davier:2019can},
taking into account all
relevant data for the electroweak (EW) sector of the `$\mu$~from~$\nu$' Supersymmetric Standard Model
(\mnSSM)~\cite{LopezFogliani:2005yw} (for a recent review of the model, see Ref.~\cite{Lopez-Fogliani:2020gzo}).
The experimental results imposed comprise 
(as will be detailed in \refse{sec:model-constraints}) Higgs and neutrino data, flavor observables such as 
$B$ and $\mu$ decays, as well as
the direct searches at the
LHC~\cite{Aad:2019vnb,Aaboud:2018zeb,Aaboud:2018jiw,Aad:2015rba,Aad:2019tcc,Sirunyan:2020eab,Sirunyan:2018ubx,Chatrchyan:2012jna}.
Sampling the $\mn$, 
it was analyzed which parameter (combinations) are favored by $\amu$ measurements. It was found that 
the $\mn$ can naturally
produce moderately light left-handed muon-sneutrinos ($120 \gev\lsim m_{\widetilde{\nu}_\mu}\lsim 620 \gev$) and wino-like charginos
($200 \gev \lsim m_{\widetilde{W}^\pm} \lsim 930 \gev$),
accommodating the discrepancy between experimental and SM values.  
A recent general analysis in the Minimal Supersymmetric Standard Model (MSSM)~\cite{Nilles:1983ge,Barbieri:1987xf,Haber:1984rc,Gunion:1984yn, Martin:1997ns} can be found in \citeres{Chakraborti:2020vjp,Chakraborti:2021kkr}.

In this work, we analyze the combination of the FNAL Run~1 
data with the previous BNL result~\cite{Abi:2021gix,Albahri:2021ixb}.
Using this improved bound we update the results
presented in \citere{Kpatcha:2019pve} and set new limits on the allowed
parameters space of the EW sector of the $\mn$, as shown in 
Figs.~\ref{SDE-msneu-M2-txt-LHC-TrisFill_lightgreen} and~\ref{RHsmuon} 
to be discussed below.
The results will be discussed in the context of searches for EW particles at the LHC {and future colliders}.
A recent general analysis of the impact of the new FNAL result in the
MSSM can be found in \citere{Chakraborti:2021dli}.


\section{The model and the experimental constraints}
\label{sec:model-constraints}

\subsection{The $\mu\nu$SSM and $a_{\mu}$}
\label{sec:model}


In the $\mn$~\cite{LopezFogliani:2005yw,Lopez-Fogliani:2020gzo}, the particle content of the MSSM
is extended by right-handed neutrino superfields $\hat \nu^c_i$. 
{The simplest superpotential of the model 
is the following~\cite{LopezFogliani:2005yw,Escudero:2008jg,Ghosh:2017yeh}: 
\bea
W &=&
\epsilon_{ab} \left(
Y_{e_{ij}}
\, \hat H_d^a\, \hat L^b_i \, \hat e_j^c +
Y_{d_{ij}} 
\, 
\hat H_d^a\, \hat Q^{b}_{i} \, \hat d_{j}^{c} 
+
Y_{u_{ij}} 
\, 
\hat H_u^b\, \hat Q^{a}
\, \hat u_{j}^{c}
\right)
\nonumber\\
&+& 
\epsilon_{ab} \left(
Y_{{\nu}_{ij}} 
\, \hat H_u^b\, \hat L^a_i \, \hat \nu^c_j
-
\lambda_i \, \hat \nu^c_i\, \hat H_u^b \hat H_d^a
\right)
+
\frac{1}{3}
\kappa_{ijk}
\hat \nu^c_i\hat \nu^c_j\hat \nu^c_k,
\label{superpotential}
\eea
where the summation convention is implied on repeated indices, with $i,j,k=1,2,3$ the usual family indices of the SM 
and $a,b=1,2$ $SU(2)_L$ indices with $\epsilon_{ab}$ the totally antisymmetric tensor, $\epsilon_{12}= 1$. 
}

{Working in the framework of a typical low-energy supersymmetry (SUSY), the Lagrangian  containing the soft SUSY-breaking terms related to $W$ 
is given by:
\bea
-\mathcal{L}_{\text{soft}}  =&&
\epsilon_{ab} \left(
T_{e_{ij}} \, H_d^a  \, \widetilde L^b_{iL}  \, \widetilde e_{jR}^* +
T_{d_{ij}} \, H_d^a\,   \widetilde Q^b_{iL} \, \widetilde d_{jR}^{*} 
+
T_{u_{ij}} \,  H_u^b \widetilde Q^a_{iL} \widetilde u_{jR}^*
+ \text{h.c.}
\right)
\nonumber \\
&+&
\epsilon_{ab} \left(
T_{{\nu}_{ij}} \, H_u^b \, \widetilde L^a_{iL} \widetilde \nu_{jR}^*
- 
T_{{\lambda}_{i}} \, \widetilde \nu_{iR}^*
\, H_d^a  H_u^b
+ \frac{1}{3} T_{{\kappa}_{ijk}} \, \widetilde \nu_{iR}^*
\widetilde \nu_{jR}^*
\widetilde \nu_{kR}^*
\
+ \text{h.c.}\right)
\nonumber\\
&+&   
m_{\widetilde{Q}_{ijL}}^2
\widetilde{Q}_{iL}^{a*}
\widetilde{Q}^a_{jL}
{+}
m_{\widetilde{u}_{ijR}}^{2}
\widetilde{u}_{iR}^*
\widetilde u_{jR}
+ 
m_{\widetilde{d}_{ijR}}^2
\widetilde{d}_{iR}^*
\widetilde d_{jR}
+
m_{\widetilde{L}_{ijL}}^2
\widetilde{L}_{iL}^{a*}  
\widetilde{L}^a_{jL}
\nonumber\\
&+&
m_{\widetilde{\nu}_{ijR}}^2
\widetilde{\nu}_{iR}^*
\widetilde\nu_{jR} 
+
m_{\widetilde{e}_{ijR}}^2
\widetilde{e}_{iR}^*
\widetilde e_{jR}
+ 
m_{H_d}^2 {H^a_d}^*
H^a_d + m_{H_u}^2 {H^a_u}^*
H^a_u
\nonumber \\
&+&  \frac{1}{2}\, \left(M_3\, {\widetilde g}\, {\widetilde g}
+
M_2\, {\widetilde{W}}\, {\widetilde{W}}
+M_1\, {\widetilde B}^0 \, {\widetilde B}^0 + \text{h.c.} \right).
\label{2:Vsoft}
\eea
In case of following the usual assumption based on the breaking of supergravity, that 
all the soft trilinear parameters are proportional to their corresponding couplings
in the superpotential (for a review, see e.g. Ref.~\cite{Brignole:1997dp}), one can write
\bea
T_{{e}_{ij}} &=& A_{{e}_{ij}} Y_{{e}_{ij}}, \;\;
T_{{d}_{ij}} = A_{{d}_{ij}} Y_{{d}_{ij}}, \;\;
T_{{u}_{ij}} = A_{{u}_{ij}} Y_{{u}_{ij}},
\nonumber \\
T_{{\nu}_{ij}} &=& A_{{\nu}_{ij}} Y_{{\nu}_{ij}}, \;\;\;
T_{{\lambda}_i}= A_{{\lambda}_i}\lambda_i, \;\;\;\;
T_{{\kappa}_{ijk}}= A_{{\kappa}_{ijk}} \kappa_{ijk},
\label{tmunu}
\eea
where 
the summation convention on repeated indices does not apply.}

In the early universe not only the EW symmetry is broken, but in addition to the neutral components of the Higgs doublet fields $H_d$ and $H_u$ also the left-handed (LH) and right-handed (RH) scalar neutrinos $\widetilde\nu_{i}$ and $\widetilde\nu_{iR}$
acquire a vacuum expectation value (vev). {With the choice of CP conservation, they develop real vevs denoted by:  
\begin{eqnarray}
\langle H_{d}^0\rangle = \frac{v_{d}}{\sqrt 2},\quad 
\langle H_{u}^0\rangle = \frac{v_{u}}{\sqrt 2},\quad 
\langle \widetilde \nu_{iR}\rangle = \frac{v_{iR}}{\sqrt 2},\quad 
\langle \widetilde \nu_{i}\rangle = \frac{v_{i}}{\sqrt 2}.
\end{eqnarray}
The EW symmetry breaking is induced {by the scalar and gaugino soft SUSY-breaking masses and
soft SUSY-breaking trilinear parameters $A$, which are all
of {\order{1 \tev}},
and therefore {also}
$v_{iR}\sim {\order{1 \tev}}$ as a consequence of their minimization equations in the scalar potential~\cite{LopezFogliani:2005yw,Escudero:2008jg,Ghosh:2017yeh}.
}
Since $\widetilde\nu_{iR}$ are gauge-singlet fields,
the $\mu$-problem can be solved in total analogy to the Next-to-MSSM
(NMSSM)~\cite{Maniatis:2009re,Ellwanger:2009dp}
through the presence in the superpotential (\ref{superpotential}) of the trilinear 
terms $\lambda_{i} \, \hat \nu^c_i\,\hat H_d \hat H_u$.
Then, the value of the effective $\mu$-parameter is given by 
$\mu=
\la_i v_{iR}/\sqrt 2$.
These trilinear terms also relate the origin of the $\mu$-term to the origin of neutrino masses and mixing angles, since the Yukawa couplings $Y_{{\nu}_{ij}} \, \hat H_u\, \hat L_i \, \hat \nu^c_j$ are present in the superpotential {generating Dirac masses for neutrinos, 
$m_{{\mathcal{D}_{ij}}}\equiv Y_{{\nu}_{ij}} {v_u}/{\sqrt 2}$.}
Remarkably, in the \mnSSM\ it is possible to accommodate neutrino masses
and mixings in agreement with experiments~\cite{Capozzi:2017ipn,deSalas:2017kay,deSalas:2018bym,Esteban:2018azc}, via an EW seesaw
mechanism dynamically generated during the EW symmetry breaking~\cite{LopezFogliani:2005yw,Escudero:2008jg,Ghosh:2008yh,Bartl:2009an,Fidalgo:2009dm,Ghosh:2010zi,Liebler:2011tp}
through the couplings
$\kappa{_{ijk}} \hat \nu^c_i\hat \nu^c_j\hat \nu^c_k$ {
giving rise to effective Majorana masses for RH neutrinos,
${\mathcal M}_{ij}
= {2}\kappa_{ijk} \frac{v_{kR}}{\sqrt 2}$.
Actually, this is possible at tree level even with diagonal Yukawa couplings~\cite{Ghosh:2008yh,Fidalgo:2009dm}, i.e.\
$Y_{{\nu}_{ij}}=Y_{{\nu}_{i}}\delta_{ij}$.}

Therefore, the \mnSSM\ solves the
$\mu$- and the $\nu$-problem (neutrino masses) simultaneously without
the need to introduce additional energy scales beyond the SUSY-breaking scale. In contrast to the (N)MSSM, $R$-parity
and lepton number are not conserved,
leading to a completely different
phenomenology characterized by distinct prompt or displaced
decays of the lightest supersymmetric particle (LSP),
producing multi-leptons/jets/photons with small/moderate missing transverse energy (MET) from 
neutrinos~\cite{Ghosh:2017yeh,Lara:2018rwv,Lara:2018zvf,Kpatcha:2019gmq}.
{The low decay width of the LSP due to the smallness of neutrino masses is also related to the existence of possible candidates for decaying dark matter in the model.
This is the case of 
the gravitino~\cite{Choi:2009ng,GomezVargas:2011ph,Albert:2014hwa,GomezVargas:2017,Gomez-Vargas:2019mqk}, or the axino~\cite{Gomez-Vargas:2019vci}, with {lifetimes} greater than the age of the Universe.
It is also worth mentioning {concerning} cosmology, that baryon asymmetry might be realized in the
$\mn$ through EW baryogenesis~\cite{Chung:2010cd}. The present limit of the electric dipole moment (EDM) of the electron highly constrains the EW baryogenesis scenario (for a review see e.g. Ref.~\cite{Chupp:2017rkp}), but since in the $\mn$ there are many scalar fields, EW baryogenesis with “multi-step phase transition” may allow successful baryogenesis while evading the EDM bound. This analysis needs a dedicated study, which would be very interesting to realize in the future. 
}

The parameter space of the $\mn$, and in particular the
neutrino, neutral Higgs, slepton, chargino and neutralino sectors are 
relevant for our analysis in order to reproduce neutrino, Higgs and $a_\mu$ data.
They were discussed in Ref.~\cite{Kpatcha:2019pve}, and we refer the reader to that work for details. Here we summarize the analysis.
{Using diagonal mass matrices for the scalar fermions, in order to avoid the
strong upper bounds upon the intergenerational scalar mixing (see e.g. Ref.~\cite{Gabbiani:1996hi}), from the eight minimization conditions with respect to $v_d$, $v_u$,
$v_{iR}$ and $v_{iL}$ to facilitate the computation we prefer to eliminate
the
soft masses $m_{H_{d}}^{2}$, $m_{H_{u}}^{2}$,  
$m_{\widetilde{\nu}_{iR}}^2$ and
$m_{\widetilde{L}_{iL}}^2$
in favor
of the VEVs.
Also, we assume} for simplicity in what follows that the couplings $\lambda_i = \lambda$,
$\kappa_{ijk}=\kappa \delta_{ij}\delta_{jk}$,
and the vevs $v_{iR}= v_{R}$. Then, the higgsino mass parameter $\mu$, and Dirac and Majorana masses discussed above are given by:
\bea
\mu=3\la \frac{v_{R}}{\sqrt 2}, \;\;\;\;
m_{{\mathcal{D}_i}}= Y_{{\nu}_{i}} 
\frac{v_u}{\sqrt 2}, \;\;\;\;
{\mathcal M}
={2}\kappa \frac{v_{R}}{\sqrt 2}.
\label{mu2}    
\eea

\medskip

\noindent
For the {\it light neutrinos}, under the above assumption, one can obtain
the following simplified formula for the effective mass matrix~\cite{Fidalgo:2009dm}:
\begin{eqnarray}
\label{Limit no mixing Higgsinos gauginos}
(m_{\nu})_{ij} 
\approx
\frac{m_{{\mathcal{D}_i}} m_{{\mathcal{D}_j}} }
{3{\mathcal{M}}}
                   \left(1-3 \delta_{ij}\right)
                   -\frac{v_{i}v_{j}}
                   {4M}, \;\;\;\;\;\;\;\;
        \frac{1}{M} \equiv \frac{g'^2}{M_1} + \frac{g^2}{M_2},         
\label{neutrinoph2}
  \end{eqnarray}     
where $g'$, $g$ are the EW gauge couplings, and $M_1$, $M_2$ the bino and wino soft {SUSY-breaking masses}, respectively.
This expression arises from the generalized EW seesaw of the $\mn$, where due to $R$-parity violation (RPV) the neutral fermions have the flavor composition
$(\nu_{i},\widetilde B^0,\widetilde W^0,\widetilde H_{d}^0,\widetilde H_{u}^0,\nu_{iR})$.
{Of the three terms in Eq.~(\ref{neutrinoph2}),
the first two 
are generated through the mixing 
of $\nu_i$ with 
$\nu_{iR}$-Higgsinos, and the third one 
also include the mixing with the gauginos.
{These are the so-called $\nu_{R}$-Higgsino seesaw and gaugino seesaw, respectively~\cite{Fidalgo:2009dm}.}
}
One can see from this equation that {once ${\mathcal M}$ is fixed, as will be done in the parameter analysis of Sec.~\ref{sec:parameter},
the most crucial independent parameters determining {\it neutrino physics} are}:
\bea
Y_{\nu_i}, \, v_{i}, \, M_1, \, M_2.
\label{freeparameters}
\eea
Note that this EW scale seesaw implies $Y_{\nu_i}\lsim 10^{-6}$
driving $v_i$ to small values because of the proportional contributions to
$Y_{\nu_i}$ appearing in their minimization equations. {A rough} estimation gives
$v_i\lsim m_{{\mathcal{D}_i}}\lsim 10^{-4}$.

{Considering the normal ordering for the neutrino mass spectrum,
and taking advantage of the 
dominance of the gaugino seesaw for some of the three neutrino families,
representative solutions for neutrino physics using diagonal neutrino Yukawas were obtained in 
Ref.~\cite{Kpatcha:2019gmq}.
In particular, the so-called type 3 solutions, which have the structure
\bea
M>0, \, \text{with}\,  Y_{\nu_2} < Y_{\nu_1} < Y_{\nu_3}, \, \text{and} \, v_1<v_2\sim v_3,
\label{neutrinomassess}
\eea
\noindent are especially interesting for our analysis of $a_\mu$, since, as will be argued below, they are able to produce the LH muon-sneutrino as the lightest of all sneutrinos. 
In this case of type 3, it is easy to find solutions with the gaugino seesaw as the dominant one for the second family. Then, $v_2$ determines the corresponding neutrino mass and $Y_{\nu_2}$ can be small.
On the other hand, the normal ordering for neutrinos determines that the first family dominates the lightest mass eigenstate implying that $Y_{\nu_{1}}< Y_{\nu_{3}}$ and $v_1 < v_2,v_3$, {with both $\nu_{R}$-Higgsino and gaugino seesaws contributing significantly to the masses of the first and third family}. Taking also into account that the composition of the second and third families in the third mass eigenstate is similar, we expect $v_3 \sim v_2$. 
}

\medskip

\noindent
The {\it LH sneutrinos} are mixed with the RH sneutrinos and neutral Higgses, since the neutral scalars and pseudoscalars in the $\mn$ have the flavor composition
$(H_{d}^0, H_{u}^0, \widetilde\nu_{iR}, \widetilde\nu_{i}) $.
Nevertheless, the LH sneutrinos are basically decoupled from the other states, since
the off-diagonal terms of the mass matrix are suppressed by the small $Y_{\nu}$ and $v_{iL}$.
In addition, scalars have degenerate masses with pseudoscalars 
$m_{\widetilde{\nu}^{\mathcal{R}}_{i}}
\approx
 m_{\widetilde{\nu}^{\mathcal{I}}_{i}}
\equiv 
m_{\widetilde{\nu}_{i}}$. 
Given that $m_{\widetilde{L}_{iL}}$ is determined for the three generations from their minimization equations, as discussed above, one arrives to the following approximate tree-level expression for the three LH sneutrino masses~\cite{Escudero:2008jg,Ghosh:2008yh,Ghosh:2017yeh}:
\bea
m_{\widetilde{\nu}_{i}}^2
\approx  
\frac{
m_{{\mathcal{D}_i}}}{v_{i}}{v_{R}}
\left(\frac{-T_{{\nu}_i}}{Y_{{\nu}_i}}-
\frac{{\mathcal M}}{2}
+ \frac{\mu}{\tan\beta}
\right),
\label{evenLLL2}
\eea
%
{where we have assumed for simplicity that for all soft trilinear parameters
$T_{ij}=T_{i}\delta_{ij}$.}

{As we can see from Eq.~(\ref{evenLLL2}), the LH sneutrino masses
depend not only on LH sneutrino vevs but also on neutrino Yukawas, and as a consequence neutrino physics is very relevant for them.
For example, if we work with expression~(\ref{tmunu}) with 
$A_{{\nu}_i}$ of {\order{1 \tev}}, neutrino physics determines sneutrino masses through the prefactor
${Y_{{\nu}_i}v_u}/{v_i}$.
Thus, values of this prefactor
in the range of about $0.01-1$, i.e.
$Y_{{\nu}_i}\sim 10^{-8}-10^{-6}$, will give rise to LH sneutrino masses in the
range of about $100-1000$ GeV.
This implies that with the hierarchy of neutrino Yukawas 
$Y_{{\nu}_{2}}\sim 10^{-8}-10^{-7}<Y_{{\nu}_{1,3}}\sim 10^{-6}$, we can obtain a
 $\widetilde{\nu}_{\mu}$ with a mass around 100 GeV whereas the masses of
 $\widetilde{\nu}_{e,\tau}$ are of the order of the TeV,
 i.e. 
 we have $m_{\widetilde{\nu}_{2}}$ as the smallest of all the sneutrino masses.
 Clearly, we are in the case of solutions for neutrino physics of type 3 discussed in Eq.~(\ref{neutrinomassess}).}

{
In Fig.~\ref{S1-NuParams-vs-Yvi} from Ref.~\cite{Kpatcha:2019pve}, we show  
$\Delta m^2_{21}=m_2^2-m_1^2$
versus $Y_{\nu_{i}}$ and $v_i$ for the scan carried out in that work, using
the results for normal ordering from Ref.~\cite{Esteban:2018azc}. 
As we can see, one can obtain the hierarchy qualitatively discussed above, 
i.e. $Y_{\nu_{2}} < Y_{\nu_{1}} < Y_{\nu_{3}}$, and $v_1 < v_3\lsim v_2$.
We will carry out a similar analysis in Sec.~\ref{sec:results}, to correctly reproduce neutrino physics.
}
 
\begin{figure}[t]
  \centering
\includegraphics[width=\linewidth, height=0.38\textheight]{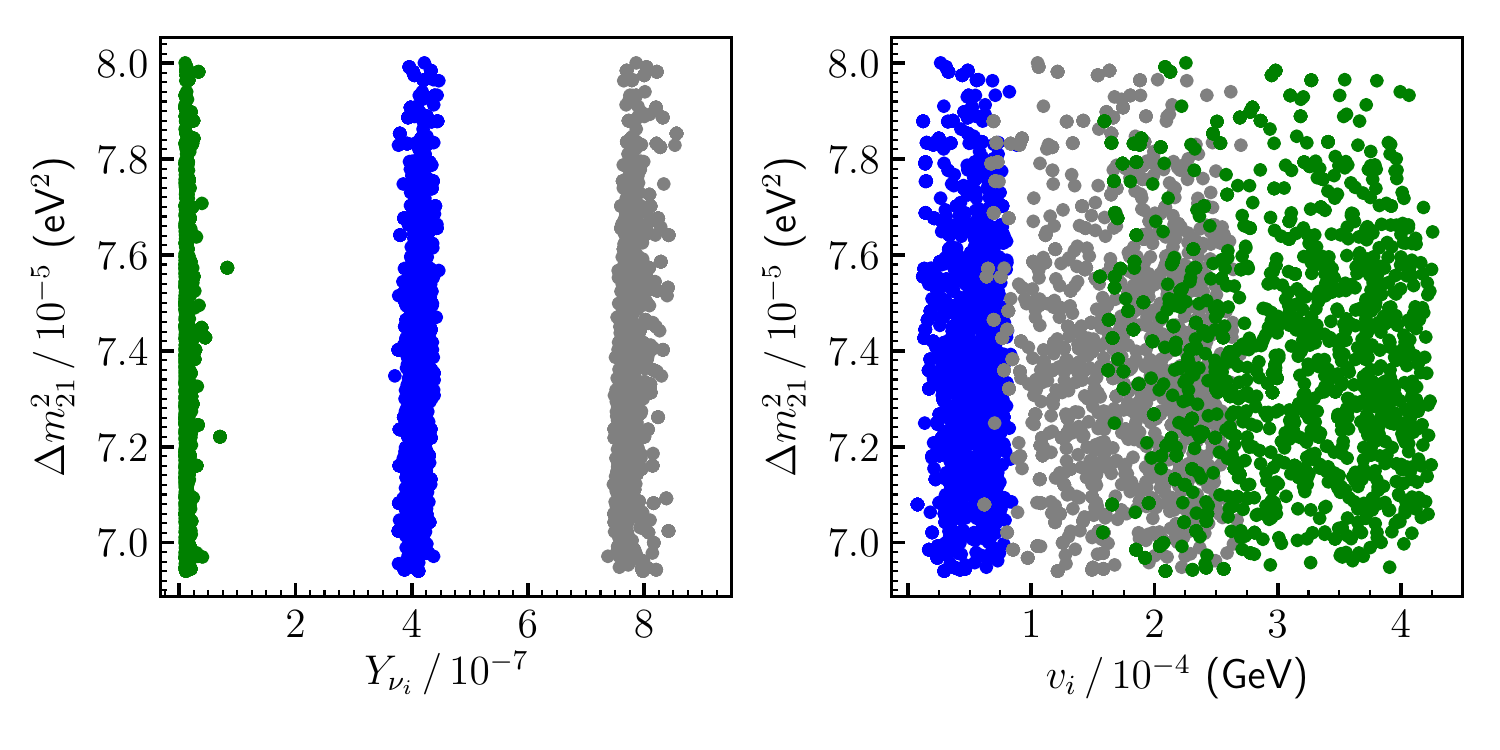}
\caption{$\Delta m^2_{21}$ versus neutrino Yukawas (left) and LH sneutrino VEVs (right).
{Colors blue, green and grey correspond to $i=1,2,3$},
 respectively.}
 \label{S1-NuParams-vs-Yvi}
\end{figure}

\medskip

\noindent
Concerning {\it charged sleptons}, in the $\mn$ LH and RH sleptons $\widetilde {\ell}_{iL,R}$ are mixed with the charged Higgses, but once again the mixing terms are suppressed and both sectors are essentially decoupled. The same comment applies to the mixing between $\widetilde {\ell}_{iL}$ and $\widetilde {\ell}_{iR}$ themselves, especially those of the first and second generation because the mixing terms are suppressed by $Y_{e_{ij}}$. In particular,
$m_{\widetilde \mu_L}$ and $m_{\widetilde \mu_R}$, which are relevant for our analysis of $a_\mu$,
are effectively determined by their soft SUSY-breaking masses 
$m_{\widetilde{L}_{2L}}$ and $m_{\widetilde e_{2R}}$.

Besides,
the masses of $\widetilde{\nu}_\mu$ and $\widetilde \mu_L$ are very similar since these two particles are in the same 
$SU(2)$ doublet, and therefore both masses are determined by 
$m_{\widetilde{L}_{2L}}$. At the end of the day,
$m_{\widetilde \mu_L}$ is only slightly larger than $m_{\widetilde{\nu}_{\mu}}$ due to the mass splitting produced by the corresponding $D$-term contributions,
$m_{\widetilde {\mu}_L}^{2} = m_{\widetilde{\nu}_{\mu}}^2
-m_W^2 \cos 2\beta$.

Therefore, 
{\it slepton physics} introduces two other independent parameters which are relevant for the value of $a_\mu$:
\bea
T_{{\nu}_i},\, m_{\widetilde e_{2R}}.
\label{tia}
\eea

\medskip

\noindent 
Unlike the LH sneutrinos, the other neutral scalars can be substantially mixed.
Neglecting this mixing between 
the doublet-like Higgses and the three RH sneutrinos, the expression of the tree-level mass of the {\it SM-like Higgs} is~\cite{Escudero:2008jg}:
\begin{eqnarray}
m_h^2 \approx 
m^2_Z \left(\cos^2 2\beta + 10.9\
{\lambda}^2 \sin^2 2\beta\right),
\end{eqnarray}
where $\tan\beta= v_u/v_d$, and $m_Z$ denotes the mass of the $Z$~boson.
Effects lowering (raising) this mass appear when the SM-like Higgs mixes with heavier (lighter) RH sneutrinos. The one-loop corrections are basically determined by 
the third-generation soft {SUSY-breaking} parameters $m_{\widetilde u_{3R}}$, $m_{\widetilde Q_{3L}}$ and $T_{u_3}$.
These three parameters together with the coupling $\lambda$ and $\tan\beta$, are the crucial ones for Higgs physics. 
The colored sector is assumed to be heavy and thus beyond the (current) reach of the LHC. This also ensures that the model can contain a scalar boson with a mass around $\sim 125 \gev$ and properties similar to the ones of the SM Higgs boson~\cite{Biekotter:2017xmf,Biekotter:2019gtq,Kpatcha:2019qsz,Biekotter:2020ehh}.

In addition, $\ka$, $v_R$ and the trilinear parameter $T_{\kappa}$ in the 
soft Lagrangian~(\ref{2:Vsoft}),
are the key ingredients to determine
the mass scale of the {\it RH sneutrinos}~\cite{Escudero:2008jg,Ghosh:2008yh}.
For example, for $\lambda\lsim 0.01$ they are basically free from any doublet admixture, and using their minimization equations in the scalar potential
the scalar and pseudoscalar masses can be approximated respectively by~\cite{Ghosh:2014ida,Ghosh:2017yeh}:
\bea
m^2_{\widetilde{\nu}^{\mathcal{R}}_{iR}} \approx   \frac{v_R}{\sqrt 2}
\left(T_{\kappa} + \frac{v_R}{\sqrt 2}\ 4\kappa^2 \right), \quad
m^2_{\widetilde{\nu}^{\mathcal{I}}_{iR}}\approx  - \frac{v_R}{\sqrt 2}\ 3T_{\kappa}.
\label{sps-approx2}
\eea

Finally, $\lambda$ and the trilinear parameter $T_{\lambda}$ 
not only contribute to these masses
for larger values of $\lambda$, but
also control the mixing between the singlet and the doublet states and hence, they contribute in determining their mass scales as discussed in detail in Ref.~\cite{Kpatcha:2019qsz}.
We conclude that the relevant low-energy parameters
in the {\it Higgs-RH sneutrino sector} are:
\bea
\lambda, \, \kappa, \, \tan\beta, \, v_R, \, T_\kappa, \, T_\lambda, \, T_{u_3}, \,  m_{\widetilde u_{3R}},
\, m_{\widetilde Q_{3L}}.
\label{freeparameterss}
\eea

\medskip

\noindent 
{\bf SUSY contributions to $a_\mu$}

\noindent
We now turn to the SUSY contributions to $a_\mu$.
The main contribution to $a_\mu$ at the one-loop level in the $\mn$,
$\amu^{\rm \mn}$, comes from the diagram involving $\widetilde{\chi}^{\pm}-\widetilde{\nu}_\mu$ loops
displayed in Fig.~\ref{one-loop-diagrams} (left),
similarly to the MSSM.
This implies that appropriately decreasing the masses of the LH muon-sneutrino $\widetilde{\nu}_\mu$ and charginos
$\widetilde{\chi}^{\pm}$ is sufficient to
lead to a significant enhancement of $a_\mu$. 
In addition, {as in the MSSM} $ a_\mu$ increases also  with increasing $\tan\beta$.
The contribution from the diagram involving $\widetilde{\chi}^{0}-\tilde \mu$ loops displayed in
Fig.~\ref{one-loop-diagrams} (right), is typically smaller
than the 
$\widetilde{\chi}^{\pm}-\widetilde{\nu}_\mu$ one
(see e.g. Refs.~\cite{Moroi:1995yh,Cerdeno:2001aj,Endo:2020mqz}). Nevertheless, we will include in our analysis variations in the masses of neutralinos $\widetilde{\chi}^{0}$ and smuons $\tilde \mu$.
In this way, $\amu^{\rm \mn}$ could have a further increase for low values of these masses.

Summarizing, for reproducing the value of
$a_\mu$
we are interested in light charginos, neutralinos, LH muon-sneutrino and smuon, and RH smuon, while the other SUSY particles can be decoupled.

    \begin{figure}
    \centering
      \includegraphics[height=3.5cm]{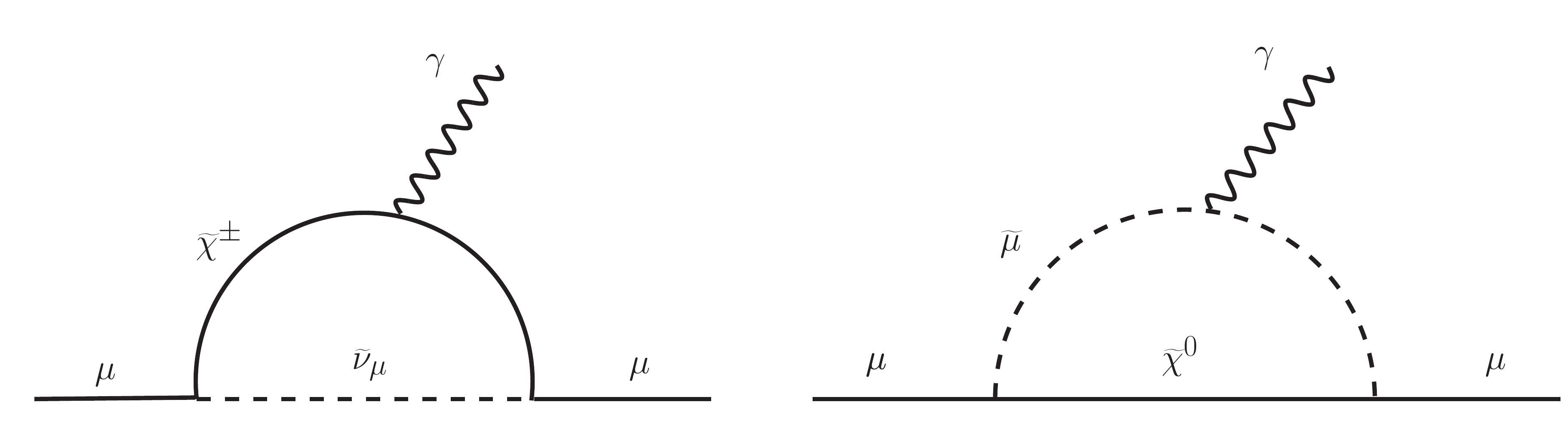}
    \caption{ Chargino-LH muon sneutrino (left) and neutralino-smuon (right) one-loop contributions to the anomalous magnetic moment of the muon.}
    \label{one-loop-diagrams}
    \end{figure}

Concerning $\widetilde{\chi}^{\pm}$, they mix with the charged leptons in the $\mn$ but the mixing terms are suppressed, and both type of particles are effectively decoupled. Therefore $\widetilde{\chi}^{\pm}$ are of the MSSM type composed of charged winos and higgsinos,
thus $m_{\widetilde{\chi}^{\pm}}$ are determined by $M_2$ and $\mu$. Note that $M_2$ is also a crucial parameter for
neutrino physics (\ref{freeparameters}), and that $\mu$ is not an independent parameter as shown in 
Eq.~(\ref{mu2}).

As discussed above, $\widetilde{\chi}^{0}$ and $\nu_{iR}$ are mixed. As we can see from Eq.~(\ref{sps-approx2}), a moderate/large negative value of 
$T_{\kappa}$ is necessary to have heavy pseudoscalar {RH} sneutrinos, but this implies that the value of $\kappa$ has to be large enough in order to avoid 
too light (even tachyonic) scalar RH sneutrinos. 
As a consequence, large Majorana masses (see Eq.~(\ref{mu2})) are generated for RH neutrinos. Therefore,
the decoupling of RH sneutrinos implies also 
heavy $\nu_{iR}$, and, in this case, $\widetilde{\chi}^{0}$ are of the MSSM type and composed of
neutral EW gauginos and higgsinos. 
Thus $m_{\widetilde{\chi}^{0}}$ are determined by $M_1$, $M_2$ and $\mu$, where the $M_1$ parameter is also crucial for neutrino physics (\ref{freeparameters}).

Concerning smuon and LH muon-sneutrino masses, 
we already discussed their masses and showed how they introduce the relevant parameters in Eq.~(\ref{tia}) for the analysis of $a_\mu$.

From this discussion, we conclude that of the parameters controlling the SUSY contributions to $\amu$, i.e.
 $M_2, M_1, \mu, m_{\widetilde{\nu}_{\mu}}, m_{\widetilde \mu_L}, m_{\widetilde e_{2R}}$ and $\tan\beta$,
only 
\bea
M_2, \,  M_1, \, m_{\widetilde e_{2R}}, \, \tan\beta,
\label{amu-params}
\eea
are independent in this scenario, and besides
$M_1$ and $M_2$ are also important for neutrino physics, and $\tan\beta$ for Higgs physics.

\medskip

\noindent
In our analysis of Sec.~\ref{sec:results}, we will sample the relevant parameter space of the $\mn$, which contains the independent parameters determining neutrino, slepton and
Higgs physics in Eqs.~(\ref{freeparameters}),~(\ref{tia}) and (\ref{freeparameterss}).
{Nevertheless, let us point out that the parameters for neutrino physics $Y_{\nu_i}$, $v_{iL}$, $M_1$ and $M_2$ are essentially decoupled from the parameters 
controlling Higgs physics.
Thus, for a suitable choice of 
$Y_{\nu_i}$, $v_{iL}$, $M_1$ and $M_2$
reproducing neutrino physics, there is still enough freedom to reproduce in addition Higgs data by playing with 
$\lambda$, $\kappa$, $v_R$, $\tan\beta$, etc., 
as shown in 
Refs.~\cite{Kpatcha:2019gmq,Kpatcha:2019pve}.
As a consequence, in Sec.~\ref{sec:results} we will not need to scan over most of the latter parameters, relaxing 
our demanding computing task.
Given the still large number of independent parameters} we have employed the 
{\tt Multinest}~\cite{Feroz:2008xx} algorithm as optimizer. To compute
the spectrum and the observables we have used SARAH~\cite{Staub:2013tta} to generate a 
{\tt SPheno}~\cite{Porod:2003um,Porod:2011nf} version for the model. 
In this way, we have evaluated the $\amu^{\rm \mn}$
contribution to 
\amu.

\subsection{Experimental constraints}
\label{sec:constr}

The $\amu$ constraint in~\refeq{gmt-diff} was applied at the {$\pm 2\,\sigma$} level. 
All other experimental constraints (except the LHC searches) are taken into account as follows:

\begin{itemize}

\item Neutrino observables\\
We have imposed the results for normal ordering from Ref.~\cite{Esteban:2018azc}, selecting points from the scan that lie within $\pm 3 \sigma$ of all neutrino observables. On the viable obtained points we have imposed the cosmological upper
bound on the sum of the masses of the light active neutrinos given
by $\sum m_{\nu_i} < 0.12$ eV~\cite{Aghanim:2018eyx}.

\item Higgs observables\\
The Higgs sector of the $\mn$ is extended with respect to the (N)MSSM.
For constraining the predictions in that sector of the model, we have interfaced 
{\tt HiggsBounds} {{v}}5.3.2~{\cite{Bechtle:2008jh,Bechtle:2011sb,Bechtle:2013wla,Bechtle:2015pma,Bechtle:2020pkv}} with {\tt Multinest}, using a 
conservative $\pm 3 \gev$ theoretical uncertainty on the SM-like Higgs boson in the $\mn$ as obtained with {\tt SPheno}. (As mentioned above, more refined calculations are possible, but would result only in shifts in the colored sector, which does not play a relevant role in our analysis.).
Also, in order to address whether a given Higgs scalar of the $\mn$ 
is in agreement with the signal observed by ATLAS and CMS, we have interfaced
{\tt HiggsSignals} {{v}}2.2.3~{\cite{Bechtle:2013xfa,Bechtle:2020uwn}} with {\tt Multinest}.
We require that the $p$-value reported by {\tt HiggsSignals} be larger than 5\%.

\item $B$~decays\\
$b \to s \gamma$ occurs in the SM at leading order through loop diagrams.
We have constrained the effects of new physics on the rate of this 
process using the average {experimental value of BR$(b \to s \gamma)$} $= (3.55 \pm 0.24) \times 10^{-4}$ provided in Ref.~\cite{Amhis:2012bh}. 
Similarly to the previous process, $B_s \to \mu^+\mu^-$ and  $B_d \to \mu^+\mu^-$ occur radiatively. We have used the combined results of LHCb and CMS~\cite{CMSandLHCbCollaborations:2013pla}, 
$ \text{BR} (B_s \to \mu^+ \mu^-) = (2.9 \pm 0.7) \times 10^{-9}$ and
$ \text{BR} (B_d \to \mu^+ \mu^-) = (3.6 \pm 1.6) \times 10^{-10}$. 
We put $\pm 3\sigma$ cuts from $b \to s \gamma$, $B_s \to \mu^+\mu^-$ and $B_d \to \mu^+\mu^-$, {as obtained with {\tt SPheno}}. {We have also checked that the values obtained are compatible with the $\pm 3 \sigma$ of the recent results from the LHCb collaboration \cite{Santimaria:2021}. 
}

\item $\mu \to e \gamma$ and $\mu \to e e e$\\
We have also included in our analysis the constraints {from 
BR$(\mu \to e\gamma) < 4.2\times 10^{-13}$~\cite{TheMEG:2016wtm}}
and BR$(\mu \to eee) < 1.0 \times 10^{-12}$~\cite{Bellgardt:1987du}, {as obtained with {\tt SPheno}}.


\item Chargino mass bound\\
In $R$-parity conserving (RPC) SUSY,
the lower bound on the lightest chargino mass of about $94 \gev$
depends on the spectrum of the model~\cite{Tanabashi:2018oca,Sirunyan:2018ubx}.
Although in the $\mn$ there is RPV and therefore this constraint does not apply automatically, we have chosen in our analysis a conservative limit of $m_{\widetilde \chi^\pm_1} > 92 \gev$.

\end{itemize}

\noindent
In addition, depending on the different masses and orderings of the lightest SUSY particles of the spectra found in our scan, we expect different signals at colliders.
Besides, depending on the values of the $\mn$ parameters, the decay of the LSP can be prompt or displaced. 
Altogether, there is a variety of possible signals arising from the regions of the parameter space analyzed, that we will able to constrain using the LHC searches of 
Refs.~\cite{Aad:2019vnb,Aad:2015rba,Aad:2020bay}.
We will discuss {these searches} in detail in the next subsection.


\subsection{LHC searches}
\label{lhc}

The different {masses and orderings of the lightest SUSY particles of the spectra} found in our analysis of the $\mn$ parameter space can be classified in four cases: 

\vspace{0.3cm}

\noindent
{\it i)} LH muon-sneutrino $\widetilde \nu_\mu$ is the LSP

\smallskip
\noindent
When $\widetilde \nu_\mu$
is the LSP 
its main decay channel corresponds to neutrinos~\cite{Ghosh:2017yeh,Lara:2018rwv,Kpatcha:2019gmq}, which constitute an invisible signal. 
Limits on sneutrino LSP from mono-jet and mono-photon searches have been discussed in the context of the $\mn$ in Refs.~\cite{Lara:2018rwv,Kpatcha:2019gmq}, and they turn out to be ineffective to constrain it. However, the presence of charginos and neutralinos in the spectrum with masses not far above {the LSP mass} is important for multi-lepton+MET searches. The relevant signals arise from the production of wino/higgsino-like chargino pairs at the LHC, which can give rise to $2\mu+4\nu$ as shown in Fig.~\ref{SneuLSP}. These processes produce a signal similar to the one expected from a directly produced pair of smuons decaying as $\widetilde{\mu}\to\mu+\widetilde{\chi}^0$ in RPC models. 
Therefore, they can be
compared with the limits obtained by the ATLAS collaboration in the search for sleptons in events with two leptons $+$ MET~\cite{Aad:2019vnb}. 

Other decay modes are possible for the wino-like charginos, in particular chains involving higgsinos when $M_2>\mu$. 
We have also considered the signals produced in events where two higgsino-like neutralinos are directly produced and decay into two smuons plus two muons, giving rise to a final signal with $4\mu +$ MET. This signal can be compared with the ATLAS search for SUSY in events with four or more leptons~\cite{Aaboud:2018zeb}. 
In this scenario, we have also considered the search for events with 2 leptons $+$ MET~\cite{Aad:2019vnb} or
3 leptons $+$ MET~\cite{Aaboud:2018jiw}, in the case where one or two of the muons would remain undetected.
As discussed in Ref.~\cite{Kpatcha:2019pve}, all these types of events cannot constrain our parameter space. 

There is a small number of points where the LH muon-sneutrino is the LSP and the RH smuon is slightly heavier. For those points, the events initiated by RH smuon pair can produce a significant number of events including leptons and missing transverse energy, however the energy of the final states is going to be too small to produce constraints.

\bigskip

\noindent
{\it ii)} Bino-like neutralino $\widetilde B^0$  is the LSP and $\widetilde \nu_\mu$ 
is the NLSP

\smallskip
\noindent
$\widetilde B^0$
can be the LSP, with $\widetilde \nu_\mu$ 
lighter than wino-like chargino/neutralino and higgsino-like chargino/neutralino and therefore the
next-to-LSP (NLSP), i.e. $m_{\widetilde{B}^0} <m_{\widetilde \nu_\mu}< m_{\widetilde W,\widetilde H}$,
where we denote both $\widetilde W^\pm$ and $\widetilde W^0$ ($\widetilde H^\pm$ and $\widetilde H^0$) generically as $\widetilde W$ ($\widetilde H$). Then, 
the proton-proton collisions produce a pair chargino-chargino, chargino-neutralino or neutralino- neutralino as shown in Fig.~\ref{binoLSP_2}. The charginos and neutralinos will rapidly decay to sneutrinos/smuons and muons/neutrinos, with the former subsequently decaying to neutrinos/muons plus binos. When
$m_{\widetilde{B}^0} \lsim m_W$ (with $m_W$ denoting the mass of the $W$~boson), $\widetilde B^0$ decay is suppressed in comparison with the one of  
$\widetilde \nu_\mu$ LSP.
This originates in the 
kinematical suppression associated with the three-body nature of the $\widetilde B^0$ decay.
For this reason, it is natural that the $\widetilde B^0$ proper decay length is an order of magnitude larger than the one of $\widetilde \nu_\mu$, being therefore of the order of ten centimeters. 
The points of the parameter space where the LSP decays with 
a proper decay 
distance larger than 1~mm can be constrained applying the limits on long-lived particles (LLPs) obtained by the ATLAS 8~TeV search~\cite{Aad:2015rba}. Let us point out that
we have used this search instead of the more 
recent 13 TeV one~\cite{Aad:2019tcc}, because it 
tests all the possible decay channels of $\widetilde B^0$ while the latter 
focuses exclusively on leptonic displaced vertices. 
The conclusion is that no points of our parameter space can be excluded by the
most recent analysis.

When $m_{\widetilde{B}^0}\gsim 130 \gev$ the two-body nature of its decay 
implies that its decay length, $c\tau$, becomes smaller than 1~mm.
In that case, following the strategy discussed in Ref.~\cite{Kpatcha:2019pve}
we can apply ATLAS searches~\cite{Aad:2019vnb}
based on the promptly produced leptons in the decay of the heavier chargino-neutralino.

{We have also considered here, and in {\it (iii)} (see below), whether the case of the direct production of a smuon pair, smuon-sneutrino or a sneutrino pair, with the smuon/sneutrino decaying into a muon/neutrino and a long lived $\widetilde B^0$, could produce a significant signal, as shown in Fig.~\ref{binoLSP_3}.
The number of events predicted in this way is added to the events produced as shown in Fig.~\ref{binoLSP_2}, and the result is that the combination of both signals excludes some points of the parameter space which are not excluded analyzing each signal separately.}

\bigskip

\noindent
{\it iii)} $\widetilde B^0$ is the LSP and 
$\widetilde W$ or 
$\widetilde H$ are co-NLSPs 

\smallskip
\noindent
The situation in this case with
$m_{\widetilde{B}^0} < m_{\widetilde{W},\widetilde H}< m_{\widetilde \nu_\mu}$
is similar to the one presented in {\it (ii)}, with the difference in the particles produced in the intermediate decay, as shown 
in Fig.~\ref{binoLSP_1}. While before, this corresponded in most cases to muons, now the intermediate decay will mainly produce hadrons.
The LHC constraints are applied in an analogous way, depending also on the value of the proper decay length, larger or smaller than 1~mm.

\bigskip

\noindent
{\it iv)} RH smuon $\widetilde \mu_{R}$ is the LSP

\smallskip
\noindent
{When $\widetilde \mu_{R}$ is the LSP
the limits from 
LEP~\cite{Heister:2002vh,Abbiendi:2005gc,Abdallah:2002rd,Abbiendi:2003ji,Abdallah:2003xe,LEPlimitsSlepton} exclude masses {smaller than 96.3~GeV for any value of its lifetime.} 
For the rest of the points with larger values of the mass
we study the following process at the LHC: once produced
$\widetilde \mu_{R}$ decays to a muon and a neutrino mediated by the small bino composition of the latter in the mass basis, as shown in Fig.~\ref{SmuonLSP}. When $\widetilde \mu_{R}$ decay is sufficiently suppressed to yield a proper decay length larger than 3~mm, it is constrained by the search for displaced leptons at ATLAS~\cite{Aad:2020bay}. 
This search is able to constrain $\widetilde \mu_{R}$ up to a mass of {500~GeV}. 
We impose the limits that are extracted {from Fig.~11d} of the auxiliary material of Ref.~\cite{1831504/t10}. 
On  the contrary, if the decay of $\widetilde \mu_{R}$ is fast enough to be considered prompt, it can be constrained using the  search for EW production of sleptons decaying into final states with two leptons and missing transverse momentum~\cite{Aad:2019vnb}. This search imposes  a {lower} limit of 450~GeV on the mass of  $\widetilde \mu_{R}$. Note that if the decay length of $\widetilde \mu_{R}$ is too long to be analyzed as prompt, but shorter than 3~mm, the LHC searches are not able to put limits on its mass.}

\begin{figure}[t!]
	\centering
	\includegraphics[
	height=0.16\textheight]{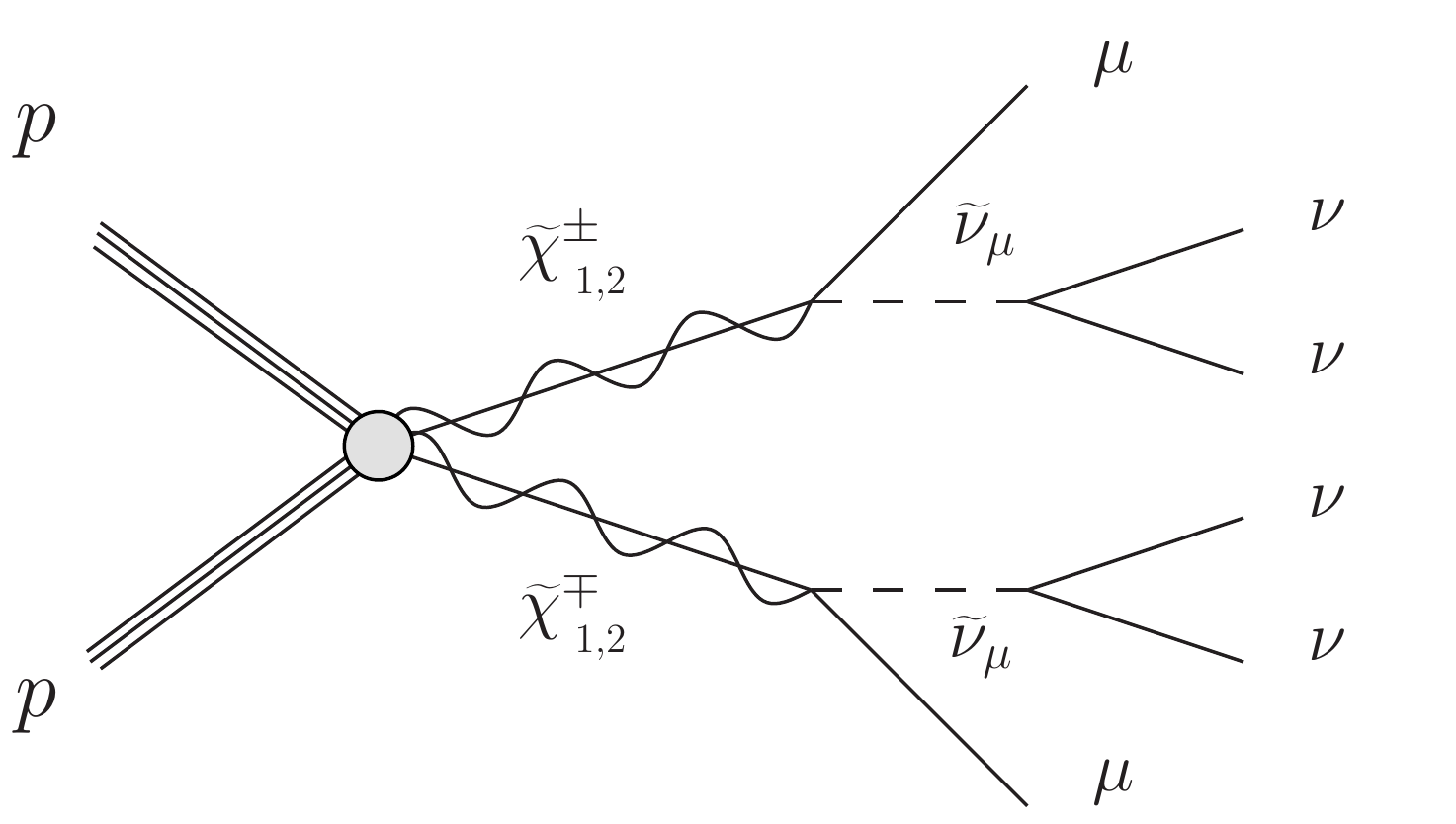}
	\caption{Production of a chargino pair, each decaying to a LH muon-sneutrino, which in turn decays to neutrinos, giving rise to the signal $2\mu+\mathrm{MET}$.
	}
	\label{SneuLSP}
\end{figure}

\begin{figure}[t!]
	\centering
	\includegraphics[
	height=0.16\textheight]{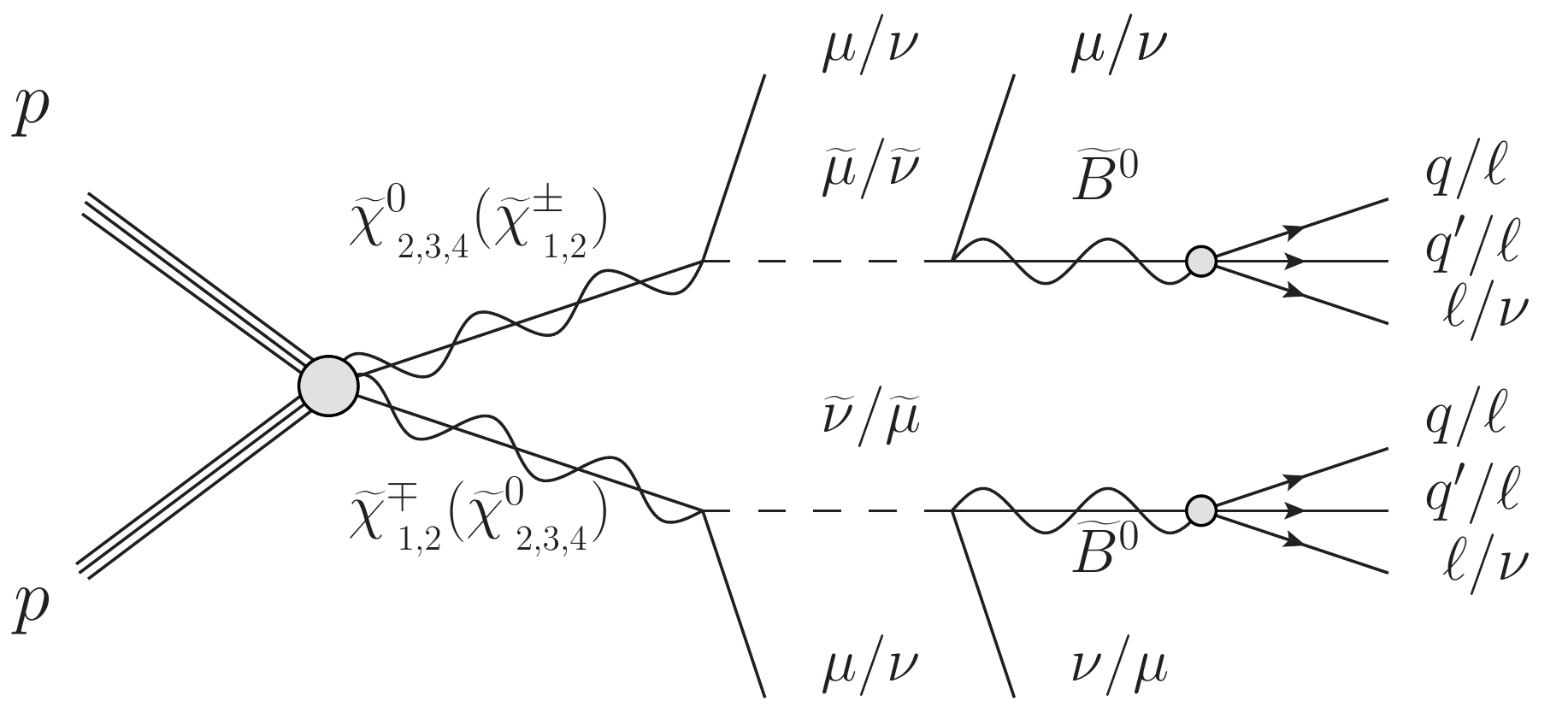}
	\caption{Production of a chargino pair, chargino-neutralino or a neutralino pair, each decaying to a LH muon-sneutrino or smuon, which in turn decay to a long-lived bino-like neutralino giving rise to a displaced signal. 
	}
	\label{binoLSP_2}
\end{figure}
\begin{figure}[t!]
	\centering
	\includegraphics[
	height=0.16\textheight]{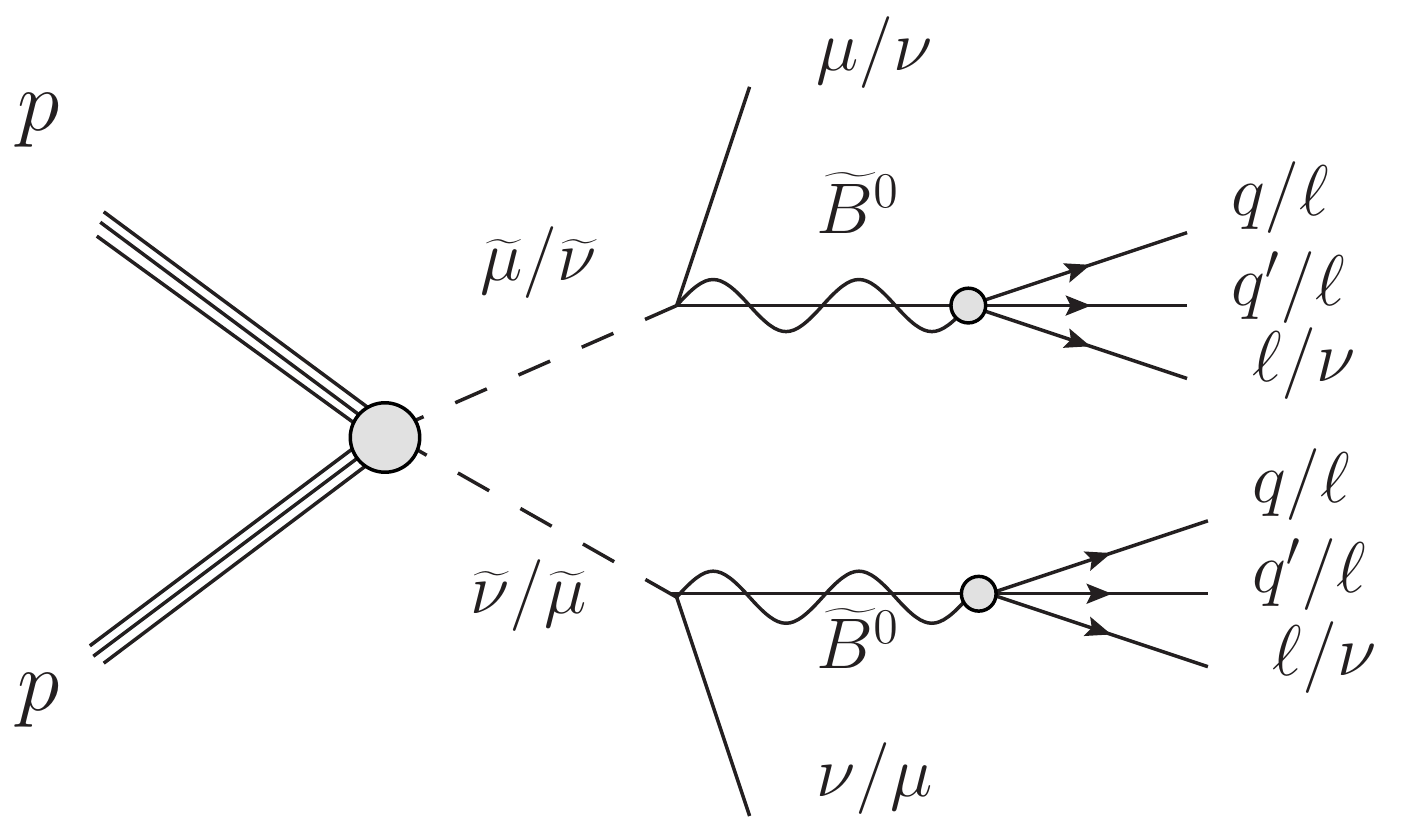}
	\caption{Production of a smuon pair, smuon-sneutrino or a sneutrino pair, each decaying to a long-lived bino-like neutralino giving rise to a displaced signal.
	}
	\label{binoLSP_3}
\end{figure}

\begin{figure}[t!]
	\centering
	\includegraphics[
	height=0.16\textheight]{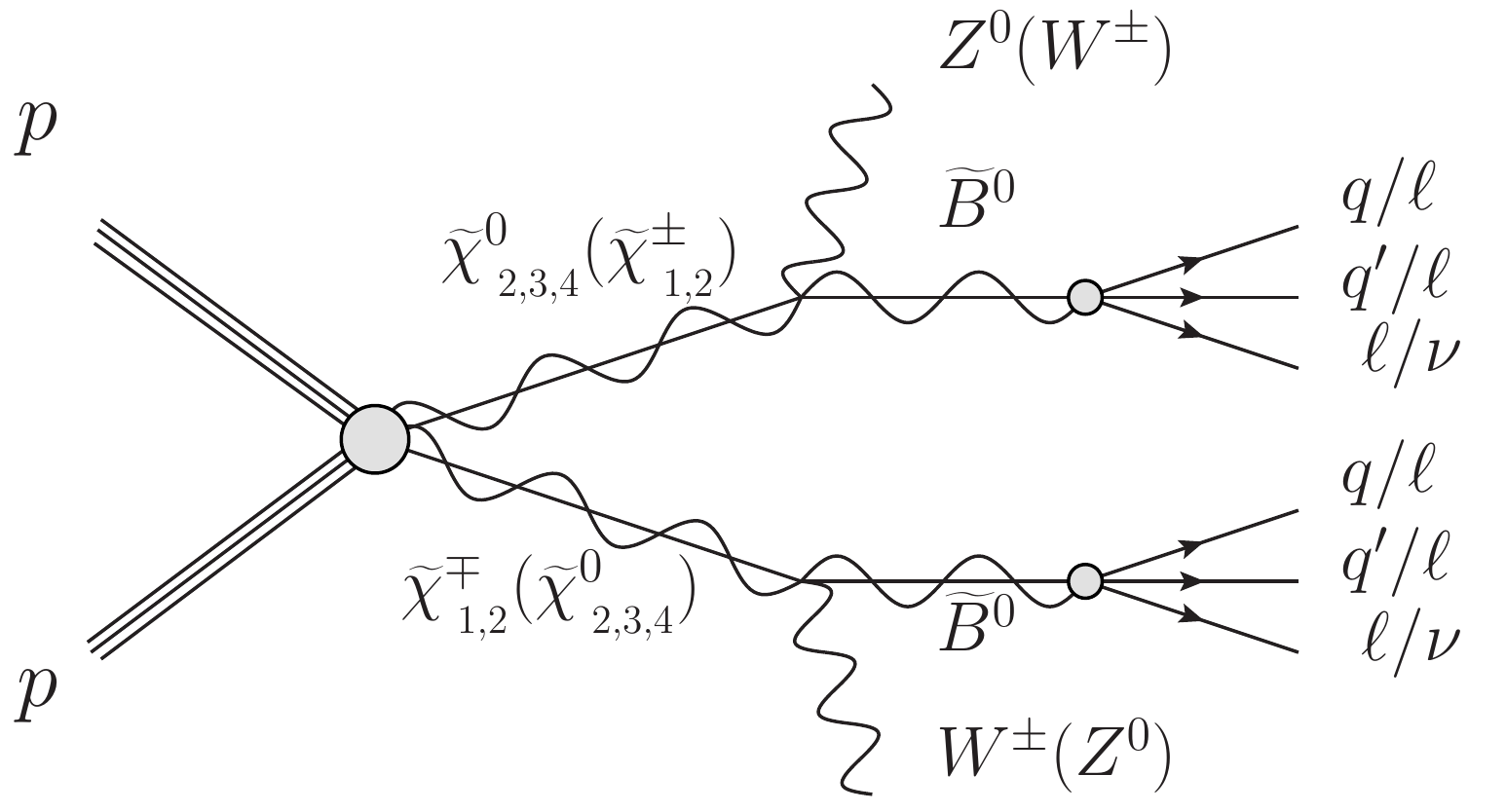}
	\caption{Production of a {chargino pair, chargino-neutralino or a neutralino pair}, each decaying to a long-lived bino-like neutralino giving rise to a displaced signal. 
	}
	\label{binoLSP_1}
\end{figure}

\begin{figure}[t!]
	\centering
	\includegraphics[
	height=0.16\textheight]{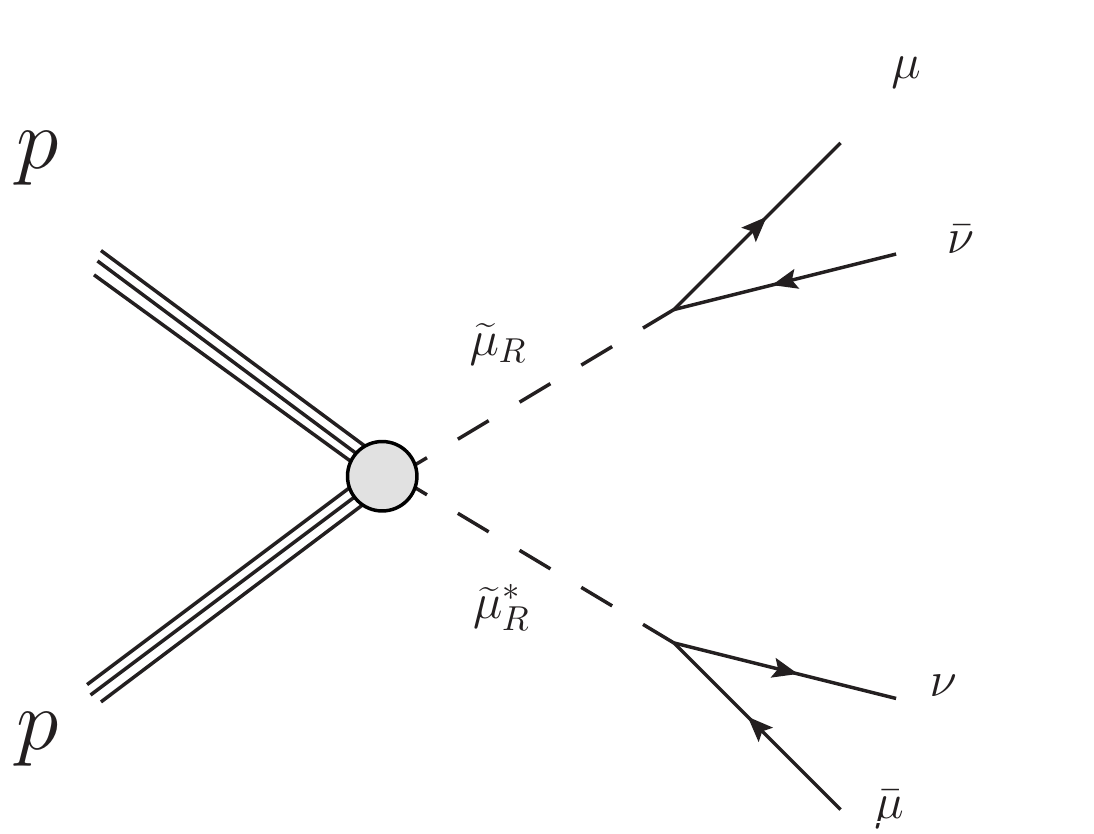}
	\caption{Production of RH smuon pair, each decaying to a RH muon and neutrino, giving rise to the signal $2\mu+\mathrm{MET}$.
	}
	\label{SmuonLSP}
\end{figure}

It is worth noting here that
when $\widetilde \mu_{R}$ is heavier than other SUSY particles, there can be three situations of interest.
In one of them, $\widetilde B^0$ is the LSP and $\widetilde \mu_{R}$ is the NLSP. 
Then, the decay of $\widetilde \mu_{R}$ will produce a signal similar to the one shown in Fig.~\ref{binoLSP_2}, but without the presence of neutrinos coming from the first step of the decay. This signal can be treated therefore in a similar way, now considering the production cross-section of a $\widetilde \mu_{R}$. 
Another situation occurs when $\widetilde{\nu}_\mu$ is the LSP and $\widetilde \mu_{R}$ the NLSP. 
The dominant decay chain in this case is $\widetilde \mu_{R} \to  W(\to l+\nu)+\widetilde{\nu}_\mu(\to \nu\nu)$, which is similar to the one analyzed in Ref.~\cite{Aad:2019vnb} (see their Fig.~1a).
Finally, we can have $\widetilde{\nu}_\mu$ as the LSP, $\widetilde B^0$ as the NLSP, and $\widetilde \mu_{R}$ slightly heavier than both of them. In this case, the dominant $\widetilde \mu_{R}$ decay chain is  $\widetilde \mu_{R} \to \widetilde B^0 
+\mu\to \nu+\widetilde{\nu}_\mu(\to \nu\nu) +\mu$, and this signal is similar to the one analyzed in Ref.~\cite{Aad:2019vnb} (see their Fig.~1b). We conclude that the limits for a $\widetilde \mu_{R}$ heavier than the LSP, imposed as described above, do not exclude any additional point.




\section{Parameter analysis}
\label{sec:parameter}

We describe in this section the methodology that we have employed to search for points 
of the parameter space of the $\mn$ that are compatible with the given experimental data. To carry out this analysis we will follow the same strategy as in Ref.~\cite{Kpatcha:2019pve}. This consists of combining a scan of the relevant parameters 
to obtain acceptable points, with a subsequent intelligent search to increase the
number of points obtained.
It is worth noting here that we will not perform any statistical interpretation of the set of points obtained, i.e. the {\tt Multinest} algorithm is just used to obtain viable points.

The Higgsino mass parameter
$\mu$ is important in our analysis,
since it is one of the parameters controlling the SUSY 
contributions to $a_\mu$ as discussed in Sec.~\ref{sec:model}. 
Besides, since higgsinos have a mass of order $\mu$, its value
has also important implications for the analysis of the LHC constraints discussed in Sec.~\ref{lhc}.
Thus, in order to have an idea of how $\amu^{\rm \mn}$ varies with $\mu$, 
it is interesting to use two representative values of this parameter in our computation. In particular, we will use a moderate value $\mu\approx 380$ GeV, and a large value $\mu\approx 800$ GeV.

\bigskip 

\noindent 
{\it 1) Moderate $\mu\approx 380$ GeV}


\noindent 
We will perform as a {\it first step} a scan, using the fewest possible parameters in order to relax our demanding computing task.
As discussed in Sec.~\ref{sec:model},
{the most crucial} parameters for neutrino 
physics~(\ref{freeparameters})
are basically decoupled from those controlling Higgs physics~(\ref{freeparameterss}).
Thus, for a suitable scan of 
$Y_{\nu_i}$, $v_{i}$, $M_1$ and $M_2$ 
reproducing neutrino physics, there is still enough freedom to reproduce in 
addition Higgs data by playing with concrete values of 
$\lambda$, $\kappa$, $\tan\beta$, $v_R$, etc.
For $\tan\beta$ we will consider a narrow range of possible values {compatible with} Higgs physics. 
On the other hand, 
LH sneutrino masses $m_{\widetilde \nu_{i}}$, introduce in addition the parameters 
$T_{{\nu}_i}$ (see Eq.~(\ref{evenLLL2})). In particular, $T_{{\nu}_2}$ is the most relevant one for our discussion of a low $m_{\widetilde \nu_{\mu}}$,
and we will scan it in an appropriate range of small values.
Since the LH sneutrinos of the other two generations can be heavier, we can fix $T_{{\nu}_{1,3}}$ to a larger value.

\begin{table}
\begin{center}
\begin{tabular}{|c||c|}
\hline
     {\bf Parameters}&  {\bf Scan}  \\   
     \hline\hline
     $\tan\beta$ & $(10, 16)$ \\
     \hline
     $Y_{\nu_{i}}$ & $(10^{-8} , 10^{-6})$  \\
     \hline
     $v_i$ & $(10^{-6} , 10^{-3})$ \\
     \hline
     $-T_{\nu_{2}}$ & $(10^{-6} , 4\times 10^{-4})$ \\
     \hline
     $M_2$ & $(150 , 1000)$  \\
    \hline 
    $\lambda$     & 0.102 \\ 
     \hline   
    $\kappa$      & 0.4 \\ 
    \hline
    $v_R$     & 1750  \\ 
    \hline
    $T_{\lambda}$   & 340  \\ 
    \hline
    $-T_{\kappa}$    & $390$ \\ 
    \hline
    $-T_{u_{3}}$  & $4140$  \\ 
    \hline
    $m_{\widetilde Q_{3L}}$  & 2950  \\ 
    \hline
    $m_{\widetilde u_{3R}}$   & 1140  \\ 
    \hline  
    $m_{\widetilde e_{2R}}$
    & \multicolumn{1}{|c|}{ 1000 } \\
      \hline
    $m_{\widetilde e_{1,3R}}$
    & \multicolumn{1}{|c|}{ 1000 } \\
    \hline
    $m_{\widetilde Q_{1,2L}}, m_{\widetilde u_{1,2R}}, m_{\widetilde d_{1.2,3R}}$  
    & \multicolumn{1}{|c|}{ 1000 } \\
    \hline
     $ M_3$  & \multicolumn{1}{|c|}{ 2700 } \\
     \hline
    $T_{u_{1,2}}$  & \multicolumn{1}{|c|}{ 0} \\
    \hline
    $T_{d_{1,2}}$, $T_{d_{3}}$  & \multicolumn{1}{|c|}{ 0, $100$ }   \\
    \hline    
    $T_{e_{1,2}}$, $T_{e_{3}}$  & \multicolumn{1}{|c|}{ 0, $40$ }  \\
    \hline
    $-T_{\nu_{1,3}}$   & \multicolumn{1}{|c|}{ $10^{-3}$ }  \\
    \hline 
    \end{tabular}  
\end{center}
   \caption{Low-energy values of the input parameters in the scan discussed in Case {\it (\it 1)} of the text. 
   Soft {SUSY-breaking} parameters and vevs 
are given in GeV.
}
     \label{Scans-parameters}
\end{table} 

{
Summarizing, {in our first step} we will perform a scan over the 9~parameters 
$Y_{\nu_i}$, $v_i$, $T_{{\nu}_2}$, $\tan\beta$, $M_2$, as shown in Table~\ref{Scans-parameters}.
Concerning
$M_1$,
we will assume for the EW gauginos $M_1=M_2/2$. This relation is
inspired by GUTs, where the low-energy result 
$M_2= (\alpha_2/\alpha_1) M_1\simeq 2 M_1$ is obtained,
with $g_2=g$ and 
$g_1=\sqrt{5/3}\ g'$.
The ranges of $v_i$ and $Y_{\nu_i}$ are natural in the context of the EW scale seesaw of the $\mn$~\cite{Kpatcha:2019gmq,Kpatcha:2019pve}.
The range
of $T_{\nu_{2}}$ is chosen to have light $\widetilde{\nu}_{\mu}$ below about 800 GeV. This is a reasonable upper bound to be able to have sizable SUSY contributions to $a_\mu$.
In the supergravity framework where
$T_{\nu_{2}}= A_{\nu_{2}} Y_{\nu_2}$, this implies
$-A_{\nu_{2}}\in$ ($1, 4\times 10^{4}$) GeV.
}

Other benchmark
parameters relevant for Higgs physics are fixed to appropriate values, as shown
in Table~\ref{Scans-parameters}.
We choose a small/moderate value for $\lambda\approx 0.1$,
thus we are in a similar situation as in the MSSM and moderate/large values of $\tan\beta$, $|T_{u_{3}}|$, and {scalar top} masses are necessary 
to obtain through loop effects the correct SM-like Higgs mass~{\cite{Biekotter:2017xmf,Biekotter:2019gtq,Kpatcha:2019qsz,Biekotter:2020ehh}}.
To avoid the chargino mass bound of RPC SUSY, 
this value of $\lambda$ forces us to choose a moderate/large value of $v_R$ to obtain a large enough value of $\mu$ (see Eq.~(\ref{mu2})).
We choose $v_R=1750$ GeV giving rise to $\mu= 378.7$ GeV.
Thus in our scan the value of $\mu$ is fixed, as discussed above.
{Another parameter controlling the SUSY contributions to $a_\mu$ is 
$m_{\widetilde e_{2R}}$ (see Eq.~(\ref{amu-params})), and we fix it in the scan to the large value 
$m_{\widetilde e_{2R}}= 1000$ GeV.}
The parameters $\ka$ and $T_{\kappa}$ are crucial to determine the mass scale of the RH sneutrinos.
We choose $T_{\kappa}=-390 \gev$ to have heavy pseudoscalar {RH} sneutrinos (of about 1190~GeV), and therefore 
the value of $\kappa$ has to be large enough in order to avoid 
too light (even tachyonic) scalar RH sneutrinos. Choosing $\kappa=0.4$, we get masses for the latter of about $700 \gev - 755 \gev$. {This value of $\kappa$ (and the one chosen above for $v_R$) also implies that the Majorana mass is fixed to ${\mathcal M}=989.9$ GeV.}
The parameter $T_{\lambda}$ is relevant to obtain the correct values of the off-diagonal terms of the mass matrix mixing the RH sneutrinos with Higgses, and we choose for its value 340~GeV.
The values of the parameters shown below the soft SUSY-breaking RH smuon mass
$m_{\widetilde e_{2R}}$
in Table~\ref{Scans-parameters}
concern slepton, squark and gluino masses, as well as quark and lepton trilinear parameters, which are not specially relevant for our analysis of $a_\mu$.
The values chosen for
$T_{\nu_{1,3}}$ are larger that the one of $T_{\nu_{2}}$, but they are natural within the supergravity framework where
$T_{\nu_{1,3}}= A_{\nu_{1,3}} Y_{\nu_{1,3}}$, since then
larger values of the Yukawa couplings are required for similar values of 
$A_{\nu_{i}}$. This allows us to reproduce the normal ordering of neutrino masses with
$Y_{\nu_2} < Y_{\nu_1} < Y_{\nu_3}$~\cite{Kpatcha:2019gmq,Kpatcha:2019pve} (see the discussion in Eq.~(\ref{neutrinomassess})). 
In a similar way, the values of $T_{d_3}$ and $T_{e_3}$ have been chosen taking into account the corresponding Yukawa couplings.



{The {\it second step} of our analysis of the $\mn$ parameter space, consists of using suitable points from the previous scan, varying appropriately some of their associated parameters in order to explore other regions, where the new points still fulfill the experimental constraints. 
In particular, note that in fact  
neutrino physics depends mainly on the parameter $M$ defined in Eq.~(\ref{neutrinoph2}).
This implies that for a given value of $M$ reproducing the correct neutrino (and Higgs) physics, one can get different pairs of values of $M_1$ and $M_2$ with the same good properties, without essentially modifying the values of the other parameters. 
This allows us to break the GUT-inspired relation $M_2=2M_1$, to explore other interesting regions of the parameter space. 
On the other hand, given the value of $m_{\widetilde \nu_{\mu}}$ 
obtained for each point, one can find more suitable points but with a different mass just varying $T_{\nu_2}$, since this parameter does not affect the neutrino (and Higgs) physics. 
In addition, 
one can also lower the soft
RH smuon mass,
{which leads to an enhancement of $\amu^{\rm \mn}$} as discussed in 
Sec.~\ref{sec:model}.
}

\bigskip 

\noindent 
{\it 2) Large $\mu\approx 800$ GeV}

\noindent 
Although the value of $\mu$ used above is reasonable,
many other values reproducing the correct Higgs physics can be obtained~\cite{Kpatcha:2019qsz}, and in particular larger ones. 
Thus we have also studied 
points with  
$\lambda= 0.126$, similar to the value above, but with a larger value for the vev $v_R=3000$ GeV giving rise to $\mu= 801.9$ GeV.
Other benchmark parameters relevant for Higgs physics have to be modified, 
such as 
$\kappa=0.36$ yielding ${\mathcal M}\approx 1527.4$~GeV, $-T_{\kappa} =150$~GeV, $T_{\lambda}=1000$~GeV,
$-T_{u_{3}}=4375$~GeV, $m_{\widetilde Q_{3L},\widetilde u_{3R}}= 2500$ GeV and
$M_3 = 3500$~GeV. For the other squarks masses and the slepton masses we use $m_{\widetilde Q_{1,2L}}, m_{\widetilde u_{1,2R}}, m_{\widetilde d_{1.2,3R}}, m_{\widetilde e_{1,2,3R}}= 1500$~GeV. We have also modified the range of $\tan\beta$ with respect to the one in Table~\ref{Scans-parameters},
using $\tan\beta \in (25, 35)$. 
Concerning the LH muon-sneutrino mass, we have slightly increased the upper limit of $-T_{\nu_{2}}$ up to $4.4\times 10^{-4}$~GeV, and to obtain slightly smaller chargino masses we have decreased the lower limit of $M_2$ up to 100~GeV. 
The rest of the parameters are taken as 
in Table~\ref{Scans-parameters}.

Unlike the previous Case {\it (1)}, instead of starting with a scan we simplify our analysis 
choosing several benchmark points with correct Higgs and neutrino physics, and apply to them the variation of relevant parameters discussed in the second step above.
In particular, for a given $M$ we vary $M_1$ and $M_2$, and for a given 
$m_{\widetilde \nu_{\mu}}$ we vary $T_{\nu_2}$.
Obviously, more acceptable points could have been obtained with a scan, but as we will see in the next section the ones obtained are sufficient to have a good idea of the interesting regions of the $\mn$ parameter space when $\mu$ is large.

\section{Results}
\label{sec:results}

We present here the results obtained following the strategy discussed in the previous section.
In particular, we use the new {combined} \gmin2\ result at the $\pm 2\sigma$ level to obtain upper (and lower) bounds on the EW SUSY masses.
We denote the points {in different regions of the parameter space, and surviving certain constraints,
with different colors and symbols as shown in Figs.~\ref{SDE-msneu-M2-txt-LHC-TrisFill_lightgreen} and~\ref{RHsmuon}.
Concerning the colors, they denote the regions where the points are} {located in the \mnSSM\ parameter space}, as well as their agreement with the new world average of $\amu^{\rm exp}$ specifying if they are in the $1\sigma$ 
or $2\sigma$ regions of $\Delta a_\mu$ in Eq.~(\ref{gmt-diff}):





\begin{itemize}

\item {Dark-green ($1\sigma$) and dark-blue ($2\sigma$)} points correspond to 
Case {\it (1)} of the previous section
with 
$m_{\widetilde e_{2R}}=$1000 GeV.

\item {Light-green ($1\sigma$) and light-blue ($2\sigma$)} points are obtained from the dark-green ones but using
$m_{\widetilde e_{2R}}=$ 100, 150, 200, 300, 500 GeV.

\item {Orange ($2\sigma$)} points
correspond to Case {\it (2)} of the previous section
with 
$m_{\widetilde e_{2R}}= $ 1500 GeV.


\end{itemize}







\noindent 
In addition, each point surviving LHC searches can be classified in the four categories discussed in Subsec.~\ref{lhc}, depending on the different relevant signals that it produces at colliders. 
We use the following symbols:

\begin{itemize}

\item[{\it i)}] Dots {($\circ$)} are points with  $\widetilde \nu_\mu$ LSP

\item[{\it ii)}] Crosses {($+$)} are points with $\widetilde{B}^0$ LSP and the hierarchy $m_{\widetilde{B}^0} <m_{\widetilde \nu_\mu}< m_{\widetilde{W},\widetilde{H}}$

\item[{\it iii)}] Triangles {($\triangle$)} are points with $\widetilde{B}^0$ LSP and the hierarchy $m_{\widetilde{B}^0} < m_{\widetilde{W},\widetilde{H}} < m_{\widetilde \nu_\mu}$

\item[{\it iv)}] Stars {($\star$)} are points with ${\widetilde \mu_{R}}$ LSP

\end{itemize}

We show in \reffi{SDE-msneu-M2-txt-LHC-TrisFill_lightgreen} the preferred parameter ranges in the $M_2-\mmuesneu$
plane. As one can see, significant regions of the parameter space of the $\mn$ can be found being
compatible with {the experimental results from neutrino and Higgs measurements}, as well as with flavor observables, {and fulfilling the constraints from the LHC Run~1 and~2, as discussed in Sec.~\ref{sec:constr} and~\ref{lhc}. In particular, they are in agreement with the new world average for $\amu^{\rm exp}$.}
{LH muon-sneutrino masses are found} in the range 
${120 \gev} \lsim m_{\widetilde{\nu}_\mu}\lsim {540 \gev}$, 
and 
${202 \gev} \lsim M_2\lsim {560 \gev}$. These values of 
$M_2$ correspond to
wino-like chargino/neutralino masses in the range 
${200 \gev\lsim m_{\widetilde W} \lsim 597 \gev}$.
The corresponding values of $M_1$ compatible with neutrino physics are smaller than $M_2$, and they are in the range
117 GeV $\lsim M_1\lsim$ 378 GeV.
The range of bino-like neutralino masses is therefore ${114 \gev} \lsim  m_{\widetilde B^0} \lsim {370 \gev}$. The values used for
$\mu \approx {380} \gev, {800} \gev$, correspond to
higgsino-like chargino/neutralino masses in the range
333 GeV $\lsim m_{\widetilde H} \lsim 878$ GeV.

In \reffi{RHsmuon}, we show as complementary information
$\amu^{\rm \mn}$ versus $M_2$.
As we can see from both figures, the smaller $M_2$ ($m_{\widetilde \nu_\mu}$) is, the {larger} $\amu^{\rm \mn}$ becomes as expected.
{One can also} see that the {orange} points with a large value of $\mu$ ($= 801.9 \gev$) do not enter in the $1\sigma$ region of $\Delta a_\mu$, unlike the {dark- and light-green} points with $\mu= 378.7 \gev$.
As mentioned above, we have also analyzed how the $\amu^{\rm \mn}$ values obtained can be enhanced lowering the value of
$m_{\widetilde \mu_{R}}$. 
In particular, we have focused on 
the {dark-green} points of the figures, which are in the
$1\sigma$ region of $\Delta a_\mu$, using instead of
$m_{\widetilde e_{2R}}=1000 \gev$ lower values.
These points are shown in light green and light blue in Figs.~\ref{SDE-msneu-M2-txt-LHC-TrisFill_lightgreen} and~\ref{RHsmuon}.
As one can see in the two figures, many {light-green} and light-blue points with an increased neutralino-smuon contribution are found with larger values 
for $\amu^{\rm \mn}$.
The maximum value of $\amu^{\rm \mn}$ can be as large as about ${36} \times 10^{-10}$ for 
$m_{\widetilde e_{2R}}= 150 \gev$. This is to be compared with the maximum value of about ${29} \times 10^{-10}$ obtained with 
$m_{\widetilde e_{2R}}=1000 \gev$ for the {dark-green} points.

\begin{figure}[t!]
\centering
\includegraphics[width=0.9\linewidth, height=0.39\textheight]{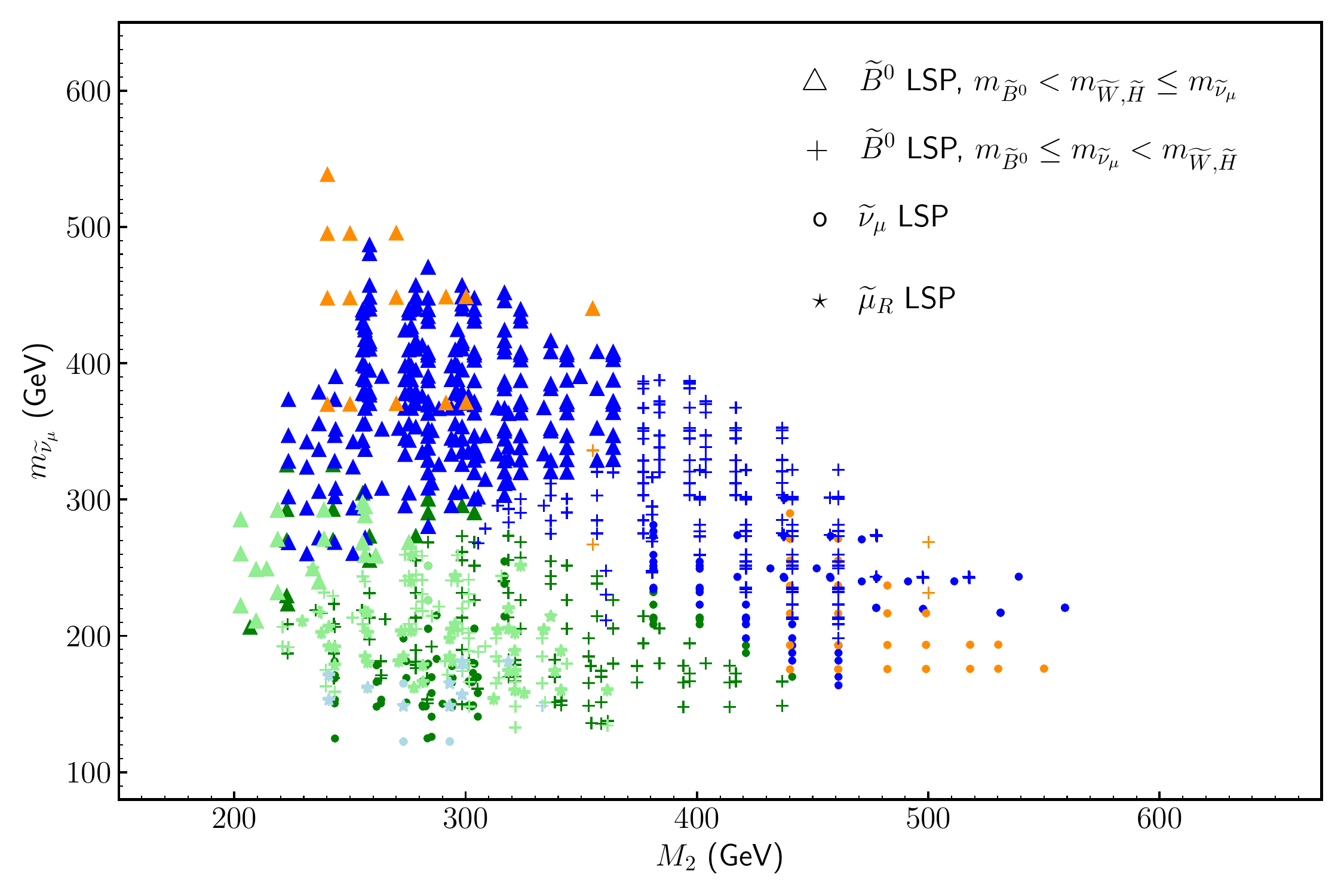}
\caption{ 
$m_{\widetilde \nu_\mu}$ versus $M_2$ for points in the parameter space of the $\mn$
in agreement with the experimental constraints.
}
 \label{SDE-msneu-M2-txt-LHC-TrisFill_lightgreen}
\end{figure}

\begin{figure}[t!]
\centering
\includegraphics[width=0.9\linewidth, height=0.39\textheight]{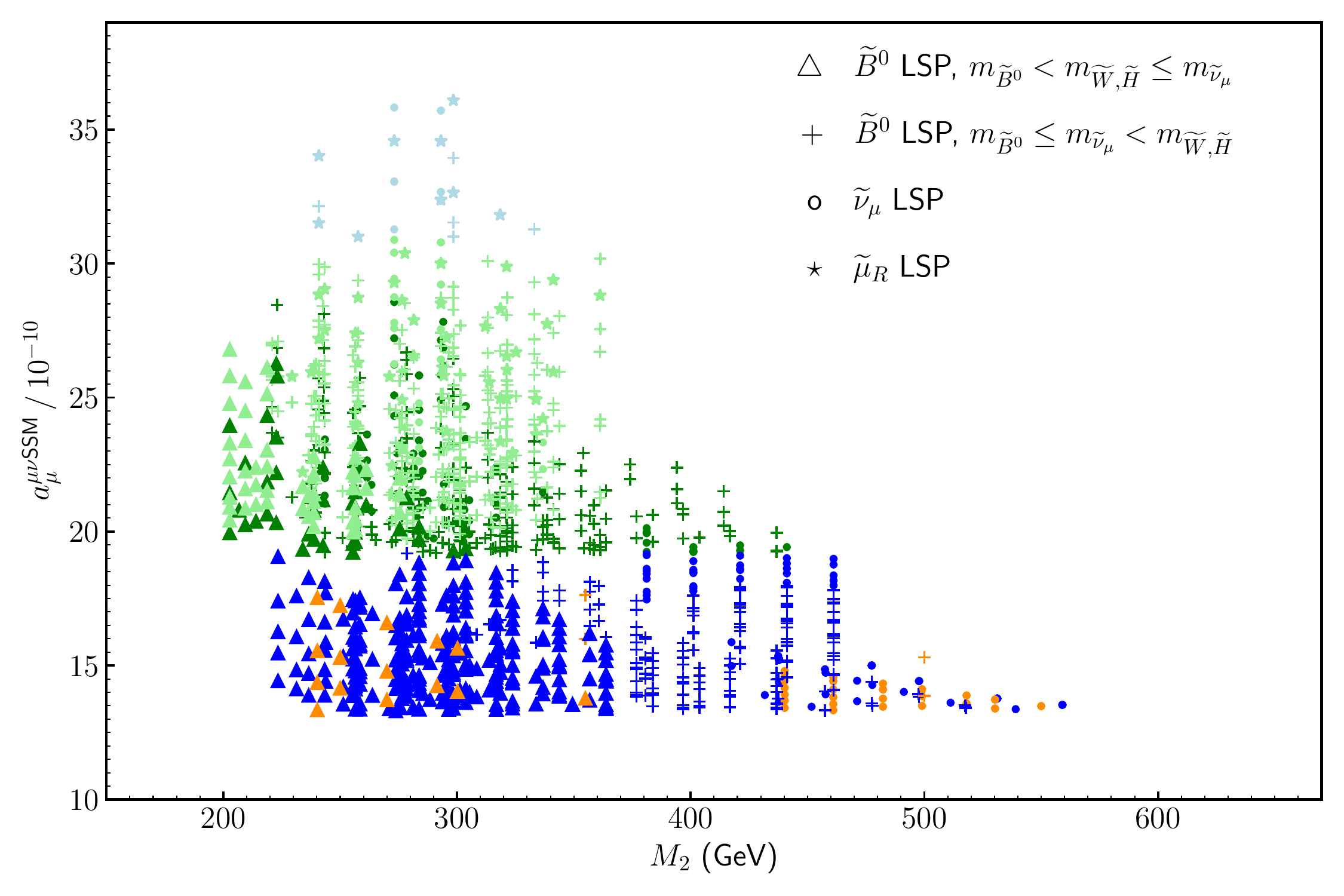}
\caption{
The same as in Fig.~\ref{SDE-msneu-M2-txt-LHC-TrisFill_lightgreen}, but showing
$\amu^{\rm \mn}$ versus $M_2$.
}
\label{RHsmuon}
\end{figure}

An interesting result of our analysis is that although LHC Runs~1 and~2 are important to constrain the $\mn$ scenario, it is not in fact difficult to find regions where many points are viable as well as compatible with $\Delta  a_{\mu}$, as shown in the figures. 
For example, a significant number of points fulfilling the experimental constrains discussed in Sec.~\ref{sec:constr} 
turn out to be forbidden by the LHC constraints discussed in Sec.~\ref{lhc}.
This is the case for many points with
$\widetilde B^0$ as the LSP, denoted by crosses and triangles. They are excluded when the GUT-inspired relation
$M_2=2 M_1$ is used. This happens for decay lengths larger as well as smaller than 1~mm, applying the displaced and prompt LHC constraints discussed in items {\it (ii)} and {\it (iii)}
of Sec.~\ref{lhc}.
Nevertheless, the situation {changes} a lot allowing $M_2\neq2 M_1$, when 
many of these points turn out to be unconstrained by LHC searches. 
{All of this kind of points in Figs.~\ref{SDE-msneu-M2-txt-LHC-TrisFill_lightgreen} and~\ref{RHsmuon} have {117 GeV $\lsim M_1\lsim 285\gev$} corresponding to a proper decay length of $\widetilde B^0$ LSP in the range
{0.1~mm $\lsim c\tau_{\widetilde B^0} \lsim$ 1.25~mm}, as shown in Fig.~\ref{SDE-ctauBino-M1_all_superimp_orange}.}

\begin{figure}[t!]
\centering
\includegraphics[width=0.7\linewidth, height=0.32\textheight]{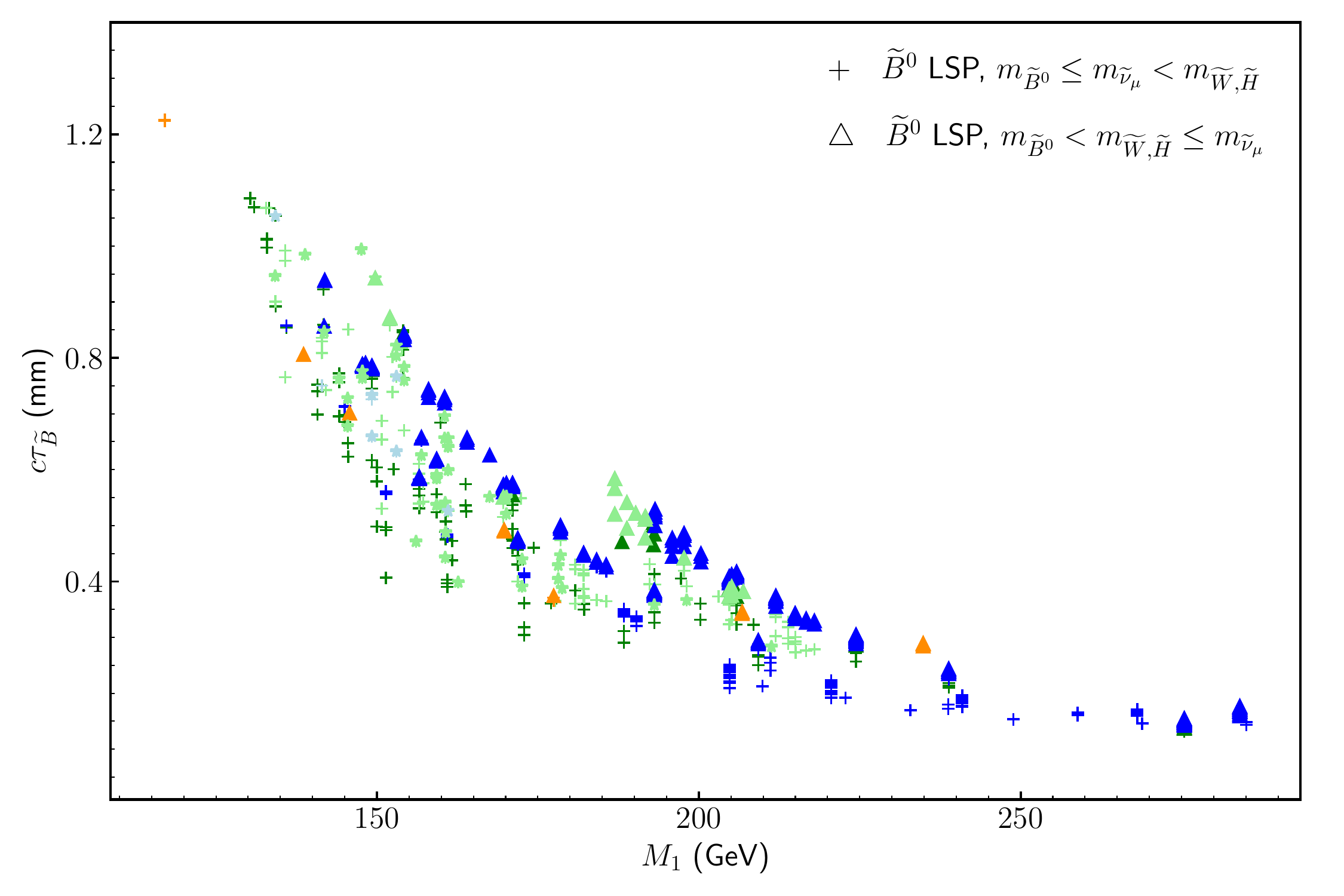}
\caption{Proper decay length of the bino LSP $c\tau_{\widetilde B^0}$ versus $M_1$.
}
 \label{SDE-ctauBino-M1_all_superimp_orange}
\end{figure}

Similarly,
the {light-green} and light-blue stars corresponding to points with $\widetilde \mu_R$ as the LSP with masses 
{$106 \gev$} $\lsim m_{\widetilde \mu_R} \lsim$ {$190 \gev$}, have a decay length in the range
{1.5~mm $\lsim c\tau_{\widetilde \mu_R} \lsim$ 3~mm}, avoiding in this way the
LEP and LHC constraints discussed in item {\it (iv)} 
of Sec.~\ref{lhc}.

Furthermore, more $\widetilde \nu_\mu$ LSP-like points are allowed when 
a larger set of $\widetilde \nu_\mu$ masses are explored  varying $T_{\nu_2}$ for given values of the rest of parameters.

It is worth noting here that a significant number of $\widetilde \nu_\mu$ LSP-like points 
are forbidden
because of the limits imposed on the higgsino-like chargino pair production. 
This can be avoided when the presence of binos with masses between the charginos and the LSP opens new decay channels. In this situation, the signal is divided into different channels that individually don't exceed the experimental limits. These points are shown in  Figs.~\ref{SDE-msneu-M2-txt-LHC-TrisFill_lightgreen} and~\ref{RHsmuon} with dots, and light- and dark-green colors
Another way to avoid it is the following. Forbidding dots can occur when $M_2>\mu$ and therefore the higgsino is lighter than the wino.
Since in our scan we fixed 
$\mu\approx 380 \gev$, points with $M_2\gsim 380 \gev$ have this hierarchy of masses.
Nevertheless, using a larger value of $\mu\approx 800 \gev$ 
allows {events initiated by higgsinos} to pass the selection cuts. 
These points are shown in Figs.~\ref{SDE-msneu-M2-txt-LHC-TrisFill_lightgreen} and~\ref{RHsmuon} with dots, and {orange colors}.

\smallskip
{Concerning the above presented analysis the following should be kept in mind.}
The values of the free parameters found {in agreement with the new} $a_\mu$ data
can be considered as a subset of all the solutions that could be obtained if a general scan of the parameter space of the model {was} carried out.
We could have obtained straightforwardly other values for
$\tan\beta$, $Y_{{\nu}_{ij}}$, 
$\lambda$, $\kappa$, $v_i$, $v_{R}$, etc., fulfilling all experimental {constraints}.
Nevertheless, we do not expect significant modifications with respect to the viable {intervals of the} values of
$m_{\widetilde \nu_\mu}$ and $M_2$ shown in
Fig.~\ref{SDE-msneu-M2-txt-LHC-TrisFill_lightgreen}, since they correspond to the sensible regions of these parameters which can give rise to a large enough {$a_{\mu}^{\mn}$}.

\section{Implications for future collider searches}
\label{sec:future3}

As we have seen in the previous section, the parameter points compatible with the {new world average for $\amu$}
predict light sleptons and/or gauginos, which can be the prime target for the future {(HL-)}LHC experiments. Generically speaking, such light {EW SUSY} particles have already been stringently restricted by the existing LHC results, especially by the multi-lepton $+$ MET searches~\cite{Aad:2019vnb, Aaboud:2018zeb, Aaboud:2018jiw}. Nevertheless, the parameter points shown above {in the $1\sigma$ region of $\Delta a_\mu$}
evade these limits thanks to {(a) metastability of the LSP such as points with triangles, crosses and stars, or (b) close mass spectrum such as points with dots. The importance of mass {degeneracy} for some of the points may be inferred from the results given in Ref.~\cite{Kpatcha:2019pve}, where both the GUT-inspired relation, $M_2 = 2 M_1$, and $M_2 \neq 2 M_1$ cases were analyzed; for $M_2 = 2 M_1$, most of the parameter points that induce a sizable value of $\amu^{\rm \mn}$ have already been excluded by the LHC experiments, while we can find many allowed points with $M_2 \neq 2 M_1$. A part of the allowed points in Figs.~\ref{SDE-msneu-M2-txt-LHC-TrisFill_lightgreen} and~\ref{RHsmuon} will be probed in the future multi-lepton $+$ MET searches at the {(HL-)}LHC, and the rest of them may be explored at a larger hadron collider such as a 100~TeV collider. For the prospects of the electroweak gaugino/slepton searches at the HL-LHC (a 100~TeV collider), see Refs.~\cite{ATLAS:2018diz, CidVidal:2018eel} (Refs.~\cite{Gori:2014oua, Arkani-Hamed:2015vfh, Golling:2016gvc}). 

Sleptons and gauginos are more efficiently probed at lepton colliders, such as the ILC~\cite{Baer:2013cma, Moortgat-Picka:2015yla, Fujii:2017vwa} and the CLIC~\cite{Moortgat-Picka:2015yla, deBlas:2018mhx}, through the pair production of these particles. For the previous studies on the role of lepton colliders in testing the SUSY explanation for $\Delta a_\mu$ in the MSSM, see Refs.~\cite{Chakraborti:2020vjp, Chakraborti:2021kkr, Endo:2013xka}. {Here usually masses up to the kinematic limit, i.e.\ $m \lsim \sqrt{s}/2$ for pair production, can be probed, see Ref.~\cite{Berggren:2020tle}. This covers compressed spectra as well as possibly meta-stable particles.} As obtained from Fig.~\ref{SDE-msneu-M2-txt-LHC-TrisFill_lightgreen}, in the present scenario, the masses of the EW SUSY particles are predicted to be {$\gtrsim 114$~GeV} and only a limited parameter space is accessible to the ILC-250. Nevertheless, most of the 1-$\sigma$ points can be probed with a higher energy, such as the ILC-500, CLIC Stage 1 (350/380 GeV), or FCC-$ee$, and the rest of the points shown in Fig.~\ref{SDE-msneu-M2-txt-LHC-TrisFill_lightgreen} may be covered at the 1-TeV ILC and CLIC Stage 2 (1.5~TeV). In addition to these electron-positron colliders, a muon collider~\cite{Delahaye:2019omf, Long:2020wfp, Ali:2021xlw} is also useful to discover new physics contributing to $\Delta a_\mu$~\cite{Capdevilla:2020qel, Buttazzo:2020eyl, Yin:2020afe, Capdevilla:2021rwo}, since the colliding muons directly couple to it. At a muon collider, not only the EW charged states---smuons, muon sneutrinos, winos, and higgsinos~\cite{Capdevilla:2021fmj}---but also binos can directly be produced through the $t$-channel smuon exchange. This direct bino production process is useful to determine the mass and lifetime of the bino in the case of the bino LSP, which then allows us to test the characteristic prediction for their correlation, as seen in Fig.~\ref{SDE-ctauBino-M1_all_superimp_orange}. We can also carry out a similar analysis for the $\widetilde{\mu}_R$ LSP.  A detailed study on the implications of future lepton colliders on the \mnSSM\,will be performed on another occasion.

In addition to the direct searches discussed above, we may probe our scenario in a more indirect manner. For example, the presence of light particles that couple to the Higgs bosons generically deviates the couplings of the SM-like Higgs boson from the SM prediction, and thus we can probe such signature through the precision measurements of the Higgs couplings.  However, for this work we chose the parameters such that to obtain a SM-like Higgs rather similar to the SM one, producing a deviation in the Higgs coupling fairly small; for instance, the change in the Higgs-photon (Higgs-muon) coupling is found to be {$\lesssim 0.8$\% (0.2\%)}, which is (far) below the sensitivity of the future precision Higgs measurements~\cite{Fujii:2017vwa}. It is, therefore, important to pursue an option for a high-energy future collider that is capable of directly producing the EW particles with a mass of {$\gtrsim 114$~GeV}.

Light sleptons and/or gauginos are also predicted in the parameter region of the MSSM where
the {new world average for $\amu$}
can be explained (for recent studies that discuss the SUSY explanation for $\Delta a_\mu$ in the MSSM based on the BNL result,  see, for instance, Refs.~\cite{Chakraborti:2020vjp, Chakraborti:2021kkr, 
%
%
%
Yanagida:2017dao, Endo:2017zrj, Hagiwara:2017lse, Bagnaschi:2017tru, Costa:2017gup, Bhattacharyya:2018inr, Gomez:2018zzw, Dutta:2018fge, Cox:2018vsv, Ibe:2019jbx, Endo:2019bcj, Badziak:2019gaf, Abdughani:2019wai, Yanagida:2020jzy, Endo:2020mqz}). Thus, a discovery of such an electroweakly charged state by itself cannot distinguish our scenario from the MSSM explanation. We {note, however,} that in the case of the MSSM, the parameter regions in which sleptons are lighter than charginos are less favored {(although not totally excluded, see Refs.~\cite{Chakraborti:2020vjp, Chakraborti:2021kkr})} by the LHC Run 2 results~\cite{Endo:2020mqz}, while such a mass spectrum is still {relatively unconstrained} in the \mnSSM. {Furthermore, in} the \mnSSM, sleptons can even be lighter than bino thanks to the $R$-parity violation, which is not allowed in the MSSM. We, therefore, envision that the determination of the mass spectrum of sleptons and gauginos may allow us to discriminate the \mnSSM\ from the MSSM. 
On the other hand, the RH sneutrinos as well as the LH sneutrinos from the first and third generation were chosen to be heavy and thus do not play a role here.

\smallskip
Displaced-vertex searches in the {LHC Run~3 or the HL-LHC} offer another promising way {to cover large parts of the favored parameter space}, since the presence of a metastable LSP is a characteristic prediction in the \mnSSM. As we discussed above, the parameter points corresponding to the cross and triangle symbols in Figs.~\ref{SDE-msneu-M2-txt-LHC-TrisFill_lightgreen} and~\ref{RHsmuon} predict a metastable bino, whose decay can be observed as a displaced decay vertex. The points shown in these figures avoid the current bound from the displaced vertex search~\cite{Aad:2015rba} since the decay length of the bino LSP is rather short, $\lesssim 1.25$~mm, as seen in Fig.~\ref{SDE-ctauBino-M1_all_superimp_orange}. We {note, however,} that it is possible to improve the sensitivity of the displaced-vertex searches for a sub-millimeter decay length; given the extremely low background in these searches, we can safely relax the requirements on the impact parameter of lepton tracks and the reconstructed position of displaced vertices, as discussed in Ref.~\cite{Lara:2018rwv}. The decay length of the bino-like neutralino LSP in our model is predicted to be larger than $0.1$~mm~\cite{Kpatcha:2019pve} (see Fig.~\ref{SDE-ctauBino-M1_all_superimp_orange}), which is sufficiently larger than the resolution of the transverse impact parameter, $\sigma (d_0)$ (for instance, $\sigma (d_0) \simeq 0.03$~mm for $p_T \gtrsim 10$~GeV~\cite{Aad:2010bx}). For the parameter points corresponding to the light-green stars, on the other hand, $\widetilde{\mu}_R$ LSP gives rise to the displaced-lepton signature. The sensitivity of the current search \cite{Aad:2020bay} is limited by the requirement on the transverse impact parameter, $|d_0|>3$~mm, which makes it insensitive to $\widetilde{\mu}_R$ with $c \tau \lesssim 3$~mm. As we mentioned above, the decay length of $\widetilde{\mu}_R$ is predicted to be in the range {$1.5~{\rm mm} \lesssim c\tau \lesssim 3 $~mm}, and thus lowering the condition on $|d_0|$ by a factor of $2-3$ may be sufficient to investigate the $\widetilde{\mu}_R$ LSP scenario. We, therefore, strongly recommend the LHC {experiments} to seriously consider the optimization of the displaced-vertex/displaced lepton searches for the sub-millimeter decay length. 

{Let us finally remark that RPV in the framework of other SUSY models has also been searched at the LHC, mainly in the MSSM. These searches typically assume either the presence of conventional trilinear lepton-number-violating couplings in the superpotential at tree level, $LLe^c$ and $LQd^c$, or the presence of trilinear baryon-number-violating couplings, $d^c d^c u^c$. 
In all these cases we expect the results concerning $a_\mu$ in the (N)MSSM to be qualitatively different from those shown in Figs.~\ref{SDE-msneu-M2-txt-LHC-TrisFill_lightgreen}--\ref{SDE-ctauBino-M1_all_superimp_orange} from superpotential~(\ref{superpotential}). The main reason is that in these plots the collider phenomenology has been taken into account, producing allowed/forbidden regions. However, the collider signals in the (N)MSSM with RPV will be generically different, as well as the corresponding dedicated LHC searches.
For example, if one introduces the $LQd^c$ or $d^c d^c u^c$ type operator, the decay of the LH muon-sneutrino is accompanied with jets, and thus its signature is totally different from that shown in Fig.~\ref{SneuLSP}. 
  If we instead have $LLe^c$, the decay of the LH muon-sneutrino gives rise to multi-lepton signature without MET, and again it is completely different from  Fig.~\ref{SneuLSP}.}






\section {Conclusions}
\label{sec:conclusion}

The EW sector of the $\mn$ can account for a variety of
experimental data, most importantly it 
can account for the long-standing discrepancy of $a_\mu$, while being
in agreement with current searches at the LHC. 
The new result for the Run~1 data of the MUON G-2 experiment
confirmed the deviation from the SM prediction found previously at BNL. 
Under the assumption that the previous experimental result on $a_\mu$ is
uncorrelated with the new MUON G-2 result, we combined the data and
obtained a new deviation from the SM prediction of
$\Delta\amu = (\newdiff \pm \newdiffunc) \times 10^{-10}$, 
corresponding to a $\newdiffsig\,\sig$ discrepancy.
We used this new result to set limits on the \mnSSM\
parameter space. 

We showed that the $\mn$ can naturally
produce light LH muon-sneutrinos and electroweak gauginos, that are consistent with
Higgs and neutrino data as well as with flavor observables such as 
$B$ and $\mu$ decays. The presence of these light sparticles
in the spectrum is known to enhance the chargino-sneutrino SUSY contribution to $a_\mu$,
and thus it is crucial for accommodating the discrepancy
between experimental and SM values.
In addition, we showed that the presence of light RH smuons increasing the neutralino-smuon contribution is helpful to obtain larger values for $a_\mu$.
We found large regions of the parameter space with these characteristics. 

We applied the constraints from LHC searches on the solutions obtained. 
The latter have a rich collider phenomenology with the possibilities of LH muon-sneutrino,  bino-like neutralino or RH smuon as LSPs.
In particular, we found that multi-lepton $+$ MET searches~\cite{Aad:2019vnb,Aad:2015rba,Aad:2020bay} can probe regions of our scenario through the production of a 
chargino pair, chargino-neutralino or a neutralino pair, as well as through the production of a smuon pair, smuon-sneutrino or a sneutrino pair, as shown in 
Figs.~\ref{SneuLSP}$-$\ref{SmuonLSP}.
Overall we found significant regions of the parameter space of the $\mn$ compatible with the {world average of $\amu$ at the $2\,\sigma$ level} and all experimental {collider} data, as shown in Figs.~\ref{SDE-msneu-M2-txt-LHC-TrisFill_lightgreen} and~\ref{RHsmuon}.
They correspond to the ranges
${202} \gev \lsim M_2 \lsim {560} \gev$,
${117} \gev \lsim M_1 \lsim {378} \gev$ and
${150} \gev \lsim m_{\widetilde e_{2R}} \lsim {1500} \gev$,
where
these values of the low-energy soft SUSY-breaking parameters 
imply that bino-like neutralino masses are in the range  
{114 GeV $\lsim m_{\widetilde B^0} \lsim 370$ GeV,}
wino-like chargino/neutralino masses 
{200 GeV $\lsim m_{\widetilde W} \lsim 597$ GeV}, and RH smuon masses {$106\lsim m_{\widetilde \mu_R} \lsim 1387$ GeV.}
The corresponding LH muon-sneutrino masses are in the range
{$120\lsim m_{\widetilde \nu_{\mu}} \lsim 540$ GeV}.
Note that the upper bounds for $M_1$ and $m_{\widetilde \mu_R}$ are an artifact of our scanning setup. Larger values of these parameters are possible, and they would only affect the small neutralino-smuon contribution to $a_\mu$.
Concerning the value of $\mu$, we used
$\mu \approx {380} \gev, {800} \gev$, corresponding to
higgsino-like chargino/neutralino masses in the range
333 GeV $\lsim m_{\widetilde H} \lsim 878$ GeV.

Finally, we discussed the implications of our results for future collider searches. As summarized in the previous paragraph, our results pin down the masses of the EW SUSY particles, and since they are predicted to be rather light, many of the allowed points in Figs.~\ref{SDE-msneu-M2-txt-LHC-TrisFill_lightgreen} and~\ref{RHsmuon} can be probed in the future multi-lepton $+$ MET searches at the (HL-)LHC. The rest of the points will be probed at a future high energy collider, such as a 100-TeV collider, the 1-TeV ILC, and the CLIC. 
Displaced-vertex searches in the future  (HL-)LHC experiments offer another promising way {to probe large parts of the favored parameter space} and to test our scenario against the MSSM explanations, since the presence of a metastable LSP is a characteristic prediction in the \mnSSM. Many points in these figures correspond to bino LSP (crosses and triangles) or RH sneutrino LSP (stars), whose decays can be observed as a displaced decay vertex or displaced leptons, respectively. To search for these metastable particles efficiently, it is important to optimize the search strategy such that a sub-millimeter displaced vertex and $\sim 1$~mm displaced lepton tracks can be detected. We highly encourage the ATLAS and CMS collaborations to consider this optimization seriously.


\begin{acknowledgments}
 
The work of S.H.\ is supported in part by the
MEINCOP Spain under contract PID2019-110058GB-C21 and in part by
the AEI through the grant IFT Centro de Excelencia Severo Ochoa SEV-2016-0597. 
The research of E.K.\ and C.M.\ was supported by the AEI
through grants 
FPA2015-65929-P (MINECO/FEDER, UE), PGC2018-095161-B-I00 and IFT Centro de Excelencia Severo Ochoa SEV-2016-0597.
The work of EK was funded by the IFT SEV-2016-0597 and Proyecto Interno UAM-125.
The work of I.L.\ was funded by the Norwegian Financial Mechanism 2014-2021, grant DEC-2019/34/H/ST2/00707. 
The work of D.L.\ was supported by the Argentinian CONICET, and also acknowledges the support through PIP 11220170100154CO. 
The work of N.N.\ was supported in part by the Grant-in-Aid for
Young Scientists (No.21K13916), Innovative Areas (No.18H05542), and Scientific Research B (No.20H01897).
S.H., E.K., I.L., C.M.\ and D.L.\ 
also acknowledge the support of the Spanish Red Consolider MultiDark FPA2017-90566-REDC.

\end{acknowledgments}

\bibliographystyle{utphys}
\bibliography{g-2-pII_munuSSM_r}

\end{document}